\documentclass[aps,prb,twocolumn,showpacs]{revtex4-1}
\usepackage{graphicx}
\usepackage{bm}
\usepackage{amsmath}

\begin{document}

\title{Spin Distribution in Diffraction Pattern of Two-dimensional Electron Gas with Spin-orbit Coupling}
\author{Cheng-Ju Lin}
\author{Chyh-Hong Chern}
\email{chchern@ntu.edu.tw} 
\affiliation{Department of Physics and Center for Theoretical Sciences, National Taiwan University, Taipei 10617, Taiwan}
\date{\today}
\begin{abstract}
Spin distribution in the diffraction pattern of two-dimensional electron gas by a split gate and a quantum point contact is computed in the presence of the spin-orbit coupling.  After diffracted, the component of spin perpendicular to the two-dimensional plane can be generated up to 0.42 $\hbar$.  The non-trivial spin distribution is the consequence of a pure spin current in the transverse direction generated by the diffraction.  The direction of the spin current can be controlled by tuning the chemical potential.
 \end{abstract}

\maketitle
\section{Introduction}

Separating spin up and spin down electrons in materials is a basic but very important procedure in spintronics.  It can be achieved by the generation of a spin current so that electrons of different spins propagate in opposite directions.  There have been many proposals of generating spin current.  They can be roughly classified into two categories.  One is by means of magnetic methods~\cite{ohno1992prl, wolf} and the other is by pure electrical sources~\cite{murakami2003scn, murakami2004prl, guo2004, guo2008}.  The first one usually acquires loops of electric current to generate magnetic field to switch electron spins.  Inevitably, heat always comes as a by-product that is notoriously unfavored,  especially when many loops are squeezed in a small volume.  Therefore, the pure electrical methods bring the ultimate hope for the sake of practical applications.

All of the electrical methods to generate spin current share the same mechanism, namely the spin-orbital interaction (SOI).  In the presence of the SOI, up spin and down spin are no longer degenerate at each point in the momentum space.  The spin orientation and momentum lock with each other in different ways for different SOI systems.  Due to the time-reversal symmetry, an electron at a particular momentum can always find its Kramer's partner at the momentum of the opposite direction.  When the SOI is very strong, some systems form topological insulators, where there is a gap in the bulk bands and there are odd numbers of the Kramer's pairs of the edge modes~\cite{kane2005, fu2007, hsieh2008}.

Although SOI plays an important role in the electrical means to control electron spin, it is the major contribution to the spin relaxation at the same time~\cite{dyakonov1972, tahan2005}.  The stronger SOI is, the shorter spin relaxation time will be.  The reason is the following.  In understanding the spin-momentum locking, one can effectively associate a fictitious magnetic field with each point in the momentum space so that the spin orientation prefers to be parallel to the fictitious magnetic field.  Suppose that an electron moves in a certain momentum, its spin is not necessary in the spin eigenstate of that momentum.  Consequently, the electron spin precesses and the information of electron spin is lost in the diffusive transport.  Unfortunately, the pros and cons of the SOI are always accompanying with one another.  Therefore, one of the most important questions in spintronics for real applications is to compromise the advantages and the disadvantages.  In this regard, a robust effect of spin splitting is needed for the spin relaxation time to be long enough for a practical purpose. 

In this paper, we propose a new effect of spin-splitting from the coherent transport by the electron diffraction.  Electron diffraction by the quantum point contact in the two-dimensional electron gas (2DEG) has been observed by the scanning probe microscopy technique.  We will further point out that there is a non-trivial spin distribution in the diffraction pattern if SOI is present.  It can be measured if the spin-resolved experimental technique is performed, for example, Kerr rotation spectroscopy. 

 The physical picture can be summarized as the following.  Because of the wave nature, electrons pick up a transverse momentum when scattered by the quantum point contact, which plays a role as a selector of propagation direction.  Because of the spin-momentum locking, electron spin passing through the quantum point contact is determined due to the selection of propagation direction.  After diffracted, the initial electron spin will certainly not be in the spin eigenstate of the diffracted momenta.  Therefore, the electron spin precesses.  Electrons in different propagation directions precess in different ways resulting in a spin distribution  in the diffraction pattern.  Different from the spin precession in the spin-relaxation mechanism, the precession in our case is coherent due to the coherent transport of the diffraction as long as there is no mixing between spin states in different bands at the same momentum.

This paper serves to provide many calculation details in our previous Letter~\cite{chern2010}.  The structure of the paper is given by the following.  The model of the theoretical proposal and the formalism will be given in the section II.  In particular, we illustrate our method using the Rashba system.  In section \ref{ssd}, we show the numerical results of the diffraction by a single slit.  In section \ref{dg}, we improve the efficiency of the spin splitting by considering the diffraction by a grating.  In section \ref{discussion}, physical origin of the new spin-splitting effect is provided  The relevancy for being realized in experiments is discussed.  In the appendix, the results for the Dresselhaus system are given.  Some formula of the special function used in the computation is included in the last appendix.

\section{The model}
The system we are considering is a 2DES with SOI diffracted by a single slit.  In experiments, a single slit may be realized by a quantum point contact.  The problem we want to solve is the spin distribution in the diffraction pattern.  The effective Hamiltonian of a 2DEG with SOI is given by the following
\begin{equation}\label{hami}
H=\frac{p^2}{2m^*}-\mu+\alpha(\hat{\sigma_x}p_y-\hat{\sigma_y}p_x)+\beta(\hat{\sigma_x}p_x-\hat{\sigma_y}p_y),\end{equation}
where $m^*$ is the effective mass of the electron, $\mu$ is the chemical potential, and $\hat{\sigma_k}$ are the Pauli spin matrices, and $\alpha$ and $\beta$ are the SOI strength corresponding to the Rashba and the Dresselhaus couplings, respectively.   Eq.~(\ref{hami}) is a $2\times 2$ Hamiltonian.  The energy bands can be obtained easily by $E^\pm_p=\frac{p^2}{2m^*}-\mu\pm\Delta_p$, where $\Delta_p=\sqrt{(\alpha^2+\beta^2)p^2+4\alpha\beta p_xp_y}$.  Apparently, SOI creates a band crossing.  $E^{+(-)}$ denotes the upper (lower) band. The eigenstates are $v_\pm=(e^{i\phi},\pm1)^T/\sqrt{2}$, where $\phi=\tan^{-1}(\frac{\alpha p_x+\beta p_y}{\beta p_x+\alpha p_y})$. For the upper band, $\phi_+=\tan^{-1}(-\frac{\alpha p_x+\beta p_y}{\beta p_x+\alpha p_y})$, and for the lower band $\phi_-=\phi_++\pi$.  $\phi$ contains the information of the spin orientation.  The orientation of the spin eigenstate is locking with the direction of momentum.  Therefore, in the system with SOI, there is no longer spin degeneracy.  Moreover, the spin orientations of the eigenstate $v_{\pm}$ lie in the $xy$ plane.  Without perturbations, the electron spin in the 2DEG has only $xy$ component.  Spin orientations are opposite for the upper band and lower band at each momentum.

The situation we consider here is similar to the case of the single slit diffraction in optics.  A slit of width $d$ is located at $y'=0$.  We compute the diffraction pattern of the electron wave after propagating to the screen at distance $L$ away in the $x$-direction.  The diffraction pattern is the superposition of the quantum waves from the slit, according to the Huygens principle.  The wave amplitude on the screen is the quantum superposition of all the spherical wave emitted from the slit.  Since the wave nature of electron is identical to photon, the diffraction pattern should be the same.  However, it is not trivial of the spin distribution in the diffraction pattern when the SOI interaction is considered.

We compute the quantum amplitude using the Green's function method pioneered by Feynman.  The quantum amplitude of the electrons from $(x', y')$ at $t=0$ to $(x,y)$ at time t is denoted by $\langle x,y,t|x',y', 0\rangle$. At $t=0$, electrons pass through the slit.  Suppose that the slit locates at $x=0$, and the screen is placed at $x=L$ away, the wave function on the screen is given by
\begin{equation}\label{wavef}
\psi(L,y,t)=\int^{d/2}_{d/2}dy'\langle L,y,t|0,y',0\rangle\phi(0,y',0),
\end{equation}
where $\phi(x', y', 0)$ is the initial wavefunction at the slit.  In the following, we compute Eq.~(\ref{wavef}) in the pure Rashba and pure Dresselhaus cases.  We shall use the pure Rashba case to illustrate our calculation procedure.  The one for the Dresselhaus case is given in the Appdendix~\ref{section1}

The pure Rashba case  is the one for $\alpha\neq0$ and $\beta = 0$.  Eq. (\ref{hami}) can be written by $H=\Lambda H_{diag}\Lambda^{\dag}$, where
\begin{eqnarray}
&&H_{diag}=\left(
\begin{array}{cc}
(\frac{p^2}{2m^*}-\mu-\alpha p)  &   0  \\
 0 &   (\frac{p^2}{2m^*}-\mu+\alpha p)   \\  
\end{array}
\right),\nonumber\\
&&\Lambda=\frac{1}{\sqrt{2}}\left(\begin{array}{cc}   ie^{-i\theta}& ie^{-i\theta}\\-1&1\end{array}\right), \nonumber\\
&&\theta=\tan^{-1}({p_y}/{p_x}).
\end{eqnarray}
The propagator in the momentum space is given by
\begin{align}\label{up}
U(t)=&e^{-\frac{i}{\hbar}Ht}=\Lambda e^{-\frac{i}{\hbar}H_{diag}t}\Lambda^\dagger \notag\\
=&\left(\begin{array}{cc}\cos(\frac{\alpha p}{\hbar}t)&e^{-i\theta}\sin(\frac{\alpha p}{\hbar}t)\\-e^{i\theta}\sin(\frac{\alpha p}{\hbar}t)&\cos(\frac{\alpha p}{\hbar}t)\end{array}\right)e^{-\frac{i}{\hbar}(\frac{p^2}{2m}-\mu)t}\notag\\=&\left(\begin{array}{cc}U_{11}&U_{12}\\U_{21}&U_{22}\end{array}\right).
\end{align}
The propagator in real space is the Fourier transformation of Eq.~(\ref{up}) giving as the following
\begin{eqnarray}\label{ur}
&\!\!&\langle x,y,t|x',y', 0\rangle\!=\!\int_{-\infty}^{\infty}\!\int_{-\infty}^{\infty}\frac{\mathrm{d}p_x\mathrm{d}p_y}{(2\pi\hbar)^2}e^{\frac{i}{\hbar}\tilde{x}p_x}e^{\frac{i}{\hbar}\tilde{y}p_y}U(t)\notag\\&\!\!&=\frac{1}{(2\pi\hbar)^2}\int_{0}^{2\pi}\!\!\!\mathrm{d}\phi_p\int_0^{\infty}\!\!p\mathrm{d}p\cdot e^{\frac{i}{\hbar}p(\tilde{x}\cos \phi_p+\tilde{y}\sin\phi_p)}U(t),
\end{eqnarray}
where $\tilde x =x-x'$, and $\tilde y = y-y'$.   Eq.~(\ref{ur}) is a $2\times 2$ matrix.  The matrix element can be computed in terms of the special functions.  In Appendix~\ref{section2}, we include the useful integration formulae of the special function.  Using Eq.(\ref{A1}) and Eq.(\ref{A2}), we obtain
\begin{eqnarray}
&&\langle x,y,t|x',y', 0\rangle_{11}=\langle x,y,t|x',y', 0\rangle_{22}\\ 
&&=\!\!\left(\frac{m}{2\pi it\hbar}\right)\!e^{\frac{i}{\hbar}\mu t}\!\sum_{n=0}^{\infty}\frac{n!}{(2n)!}\!\left(\frac{2im\alpha^2t}{\hbar}\right)^n\!\!\!\!\times\!\!\!\ _1F_1(\! n+1;\!1;\!\frac{im\tilde r^2}{2\hbar t}),\nonumber
\end{eqnarray}
where $_1F_1(a;b;z)$ is the hypergeometric function and $\tilde r^2=\tilde x^2 + \tilde y^2$.  Using Eq. (\ref{A3}), Eq. (\ref{A4}), and Eq. (\ref{A5}), we can compute the other elements given by
\begin{align}
&\langle x,y,t|x',y', 0\rangle_{12}\notag\\
&=\left(\frac{m}{2\pi it\hbar}\right)\left(\frac{m\alpha}{\hbar}\right)e^{\frac{i}{\hbar}\mu t}(\tilde x-i\tilde y)\notag\\
&\times\sum_{n=0}^{\infty}\frac{(n+1)!}{(2n+1)!}\left(\frac{2im\alpha^2t}{\hbar}\right)^n\ _1F_1(n+2;2;\frac{im\tilde r^2}{2\hbar t}\!)
\end{align}
Similarly,
\begin{align}
&\langle x,y,t|x',y', 0\rangle_{21}\notag\\
&=-\left(\frac{m}{2\pi it\hbar}\right)\left(\frac{m\alpha}{\hbar}\right)e^{\frac{i}{\hbar}\mu t}(\tilde x+i\tilde y)\notag\\
&\times\sum_{n=0}^{\infty}\frac{(n+1)!}{(2n+1)!}\left(\frac{2im\alpha^2t}{\hbar}\right)^n\ _1F_1(n+2;2;\frac{im\tilde r^2}{2\hbar t}).
\end{align}

The propagation time $t$ is determined by  $t=mL/p_{Fx}$, where $p_{Fx}$ is the $x$-component of the momentum of the electron at the Fermi energy.  For small angle diffraction as usually considered in optics, it is a good approximation.  Furthermore, we introduce the following dimensionless quantities for computational convenience.   We define $\bar \alpha \equiv m\alpha L/\hbar$ and $\bar \beta \equiv m\beta L/\hbar$, which are the dimensionless strength of the couplings, $\bar d=d/L$ and $\bar y=y/L$, and $k \equiv p_{Fx}d/2\hbar$.

The computation can be further simplified as the following.  There are four terms to compute in the matrix product in Eq.~(\ref{wavef}).  In order to make the numerical results more convergent, we express the hypergeometric functions in the integral form.  Given the initial wavefunction $\phi^-(0,y',0)=(\phi_1,\phi_2)=(i,-1)^T/\sqrt{2d}$ from the lower Rashba band, we obtain  
\begin{align}\label{psi11}
&\int^{d/2}_{-d/2}\langle x,y,t|x',y', 0\rangle_{11}\phi_1(0,y',0)\notag\\
&=\frac{i}{\sqrt{2d}}\left(\frac{m}{2\pi it\hbar}\right)e^{\frac{i}{\hbar}\mu t}\sum_{n=0}^{\infty}\frac{n!}{(2n)!}\left(\frac{i\bar \alpha^2\bar d}{k}\right)^n\notag\\
&\times\frac{1}{2\pi i}\oint\mathrm ds(s-1)^{-(n+1)}s^{n}\int_{-d/2}^{d/2}\mathrm dy'\exp(s\frac{im\tilde r^2}{2\hbar t})  \notag\\
&=\frac{i}{\sqrt{2d}}\left(\frac{md}{2\pi it\hbar}\right)e^{\frac{i}{\hbar}\mu t}F(y),
\end{align}
where $\tilde r^2 =x^2+y^2-2yy'+y'^2\approx r^2-2yy'$ and $r^2=x^2+y^2$ in the $y'$ integration are used, and $F(y)$ and $f_n(y)$ are given by
\begin{eqnarray}\label{psi12}
&&F(y)=\sum_{n=0}^{\infty}\frac{1}{(2n)!}\left(\frac{i\bar \alpha^2\bar d}{k}\right)^nf_n(y),\\
&&f_n(y)=\left[\left(\frac{d}{ds}\right)^ns^{n-1}\exp(s\frac{imr^2}{2\hbar t})\frac{\sin(sk\bar y)}{k \bar y}\right]_{s=1}. \nonumber
\end{eqnarray}
Similarly,
\begin{eqnarray}\label{psi21}
&\!\!\!&\int^{d/2}_{-d/2}\langle L,y,t|x',y', 0\rangle_{12}\phi_2(0,y',0)\nonumber\\
&\!\!\!&=\frac{-1}{\sqrt{2d}}\left(\frac{md}{2\pi it\hbar}\right)e^{\frac{i}{\hbar}\mu t}\bar \alpha\left[(1\!-\!i\bar y)G(y)\!-\!\bar dH(y)\right],
\end{eqnarray}
where
\begin{eqnarray}
&\!\!&G(y)=\sum_{n=0}^{\infty}\frac{1}{(2n+1)!}\left(\frac{i\bar \alpha^2\bar d}{k}\right)^ng_n(y),\nonumber\\
&\!\!&g_n(y)=\left[\left(\frac{d}{ds}\right)^ns^{n}\exp(s\frac{imr^2}{2\hbar t})\frac{\sin(sk\bar y)}{k \bar y}\right]_{s=1},\nonumber\\
&\!\!&H(y)=\sum_{n=0}^{\infty}\frac{1}{(2n+1)!}\left(\frac{i\bar \alpha^2\bar d}{k}\right)^nh_n(y),\nonumber\\
&\!\!&h_n(y)\!=\!\!\left[\!\!\left(\frac{d}{ds}\right)^n\!\!\!s^{n\!-\!1}\!e^{s\frac{imr^2}{2\hbar t}}\frac{sk\bar y\cos(\!sk\bar y\!)\!-\!\sin(\!sk\bar y\!)}{2k^2 \bar y^2}\right]_{\!s\!=\!1}.\nonumber
\end{eqnarray}
The other two terms can be also simplified in the same way.
\begin{eqnarray}\label{psi22}
&\!\!\!&\int^{d/2}_{-d/2}\langle x,y,t|x',y', 0\rangle_{22}\phi_2(0,y',0)\nonumber\\
&\!\!\!&=\frac{i}{\sqrt{2d}}\left(\frac{md}{2\pi it\hbar}\right)e^{\frac{i}{\hbar}\mu t}F(y) \nonumber \\ \nonumber &&\\
&\!\!\!&\int^{d/2}_{-d/2}\langle L,y,t|x',y', 0\rangle_{21}\phi_1(0,y',0)\nonumber\\
&\!\!\!&=\frac{-i}{\sqrt{2d}}\left(\frac{md}{2\pi it\hbar}\right)e^{\frac{i}{\hbar}\mu t}\bar \alpha\left[(1+i\bar y)G(y)+\bar dH(y)\right]
\end{eqnarray}
Combining Eq.~(\ref{psi11}), Eq.~(\ref{psi12}), Eq.~(\ref{psi21}), Eq.~(\ref{psi22}), the wave function $\psi^-=(\psi_1^{(-)},\psi_2^{(-)})$ on the screen can be computed as the following
\begin{eqnarray}\label{spsidrashba}
&\!\!\!&\psi_1^{(-)}=A\left[iF(\bar y)-\bar \alpha(1-i\bar y)G(\bar y)+\bar \alpha \bar d H(\bar y)\right],\nonumber\\
&\!\!\!&\psi_2^{(-)}=A\left[-F(\bar y)-i\bar \alpha(1+i\bar y)G(\bar y)-i\bar \alpha \bar d H(\bar y)\right],
\end{eqnarray}
where $A=\frac{1}{\sqrt{2d}}\left(\frac{md}{2\pi it\hbar}\right)e^{\frac{i}{\hbar}\mu t}$.  The diffraction pattern for the electrons from the upper Rashba band can be also obtained in the same way.  Given the initial wave function $\phi^+(0,y',0)=(i,1)^T/\sqrt{2d}$, the wavefunction $\psi^+=(\psi^{(+)}_1,\psi^{(+)}_2)$ on the screen is given by
\begin{eqnarray}\label{spsiurashba}
&\!\!\!&\psi_1^{(+)}=A\left[iF(\bar y)+\bar \alpha(1-i\bar y)G(\bar y)-\bar \alpha \bar d H(\bar y)\right],\nonumber\\
&\!\!\!&\psi_2^{(+)}=A\left[F(\bar y)-i\bar \alpha(1+i\bar y)G(\bar y)-i\bar \alpha \bar d H(\bar y)\right],
\end{eqnarray}

We will compute Eq.~(\ref{spsidrashba}) and Eq.~(\ref{spsiurashba}) numerically in the next section.  $|\psi|^2=|\psi_1|^2+|\psi_2|^2$ is probability distribution of electron on the screen, which is nothing but the diffraction pattern.  As $\psi_1$ and $\psi_2$ are the spin components, we will show later that the spatial distribution of $|\psi_1|^2$ and $|\psi_2|^2$ on the screen are different in Eq.~(\ref{spsidrashba}) and Eq.~(\ref{spsiurashba}).  The difference implies a nontrivial spin distribution in the diffraction pattern that we hope to investigate in this paper.

\begin{figure}[htbp] 
   \centering
   \includegraphics[scale=0.36]{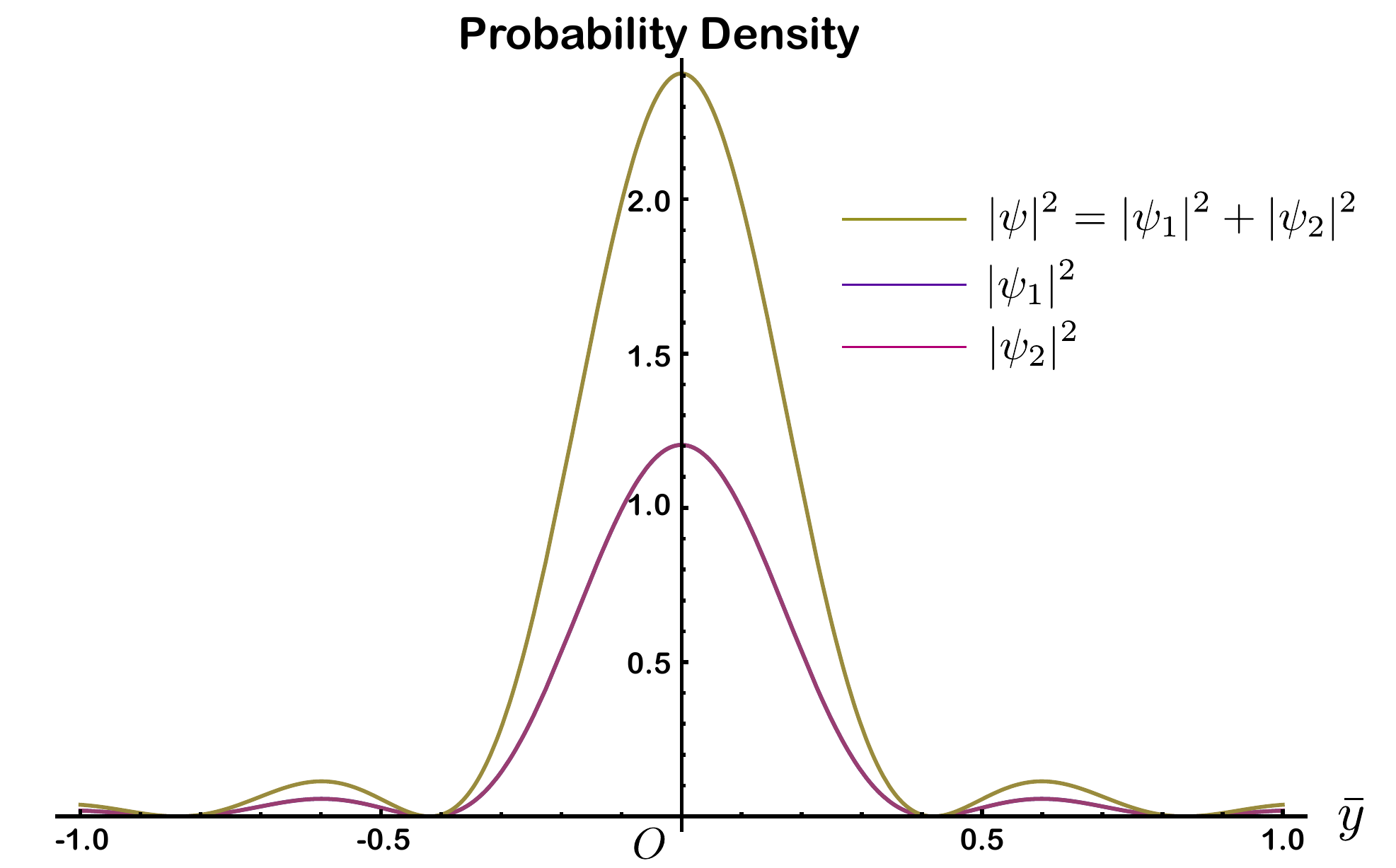} 
   \caption{(color online) The probability density of $|\psi |^2$ ,$|\psi_1|^2$ and $|\psi_2|^2$ in the dimensionless parameters $\bar{\alpha}=0, \bar{\beta}=0, k=7.5$, and $\bar d=0.01$. The diffraction pattern for spin-up electron is the same as for spin-down electron. Since there is spin degeneracy in the momentum space, the total spin of the central peak is zero.}
   \label{fig:noSOI}
\end{figure}

The pure Dresselhaus case can be computed in the similar way.  The Hamiltonian now becomes $H=\Lambda H_{diag}\Lambda^{\dag}$, where
\begin{eqnarray}
&\!\!&H_{diag}=\left(
\begin{array}{cc}
(\frac{p^2}{2m^*}-\mu-\beta p)  &   0  \\
 0 &   (\frac{p^2}{2m^*}-\mu+\beta p)   \\  
\end{array}
\right),\\
&\!\!&\Lambda=\frac{1}{\sqrt{2}}\left(\begin{array}{cc}   e^{i\theta}& e^{i\theta}\\-1&1\end{array}\right), \ \ \theta=\tan^{-1}({p_y}/{p_x})\nonumber
\end{eqnarray}
Although the energy dispersion is the same as the Rashba case, the orientation of the spin eigenstates differs.  It gives rise to different spin distribution.  Using the same computational procedure, we obtain
\begin{eqnarray}\label{spsiddress}
&\!\!\!&\psi_1^{(-)}=A\left[F(\bar y)+\bar \beta(i-\bar y)G(\bar y)+i\bar \beta \bar d H(\bar y)\right],\nonumber\\
&\!\!\!&\psi_2^{(-)}=A\left[-F(\bar y)-\bar \beta(i+\bar y)G(\bar y)+i\bar \beta \bar d H(\bar y)\right],
\end{eqnarray}
where $\psi^{(-)}=(\psi_1^{(-)},\psi_2^{(-)})$ are the wavefunction on the screen for the lower Dresselhaus band.
\begin{eqnarray}\label{spsiudress}
&\!\!&\psi_1^{(+)}=A\left[F(\bar y)-\bar \beta(i-\bar y)G(\bar y)-i\bar \beta \bar d H(\bar y)\right],\nonumber\\
&\!\!&\psi_2^{(+)}=A\left[F(\bar y)-\bar \beta(i+\bar y)G(\bar y)+i\bar \beta \bar d H(\bar y)\right],
\end{eqnarray}
where $\psi^{(+)}=(\psi_1^{(+)},\psi_2^{(+)})$ are the wavefunction on the screen for the upper Dresselhaus band.  This completes all of our analytical results in this paper.  In the next section, we will solve them numerically and investigate the magnetic property.

\section{Numerical results}
\subsection{Single-slit diffraction}\label{ssd}

We first plot $|\psi|^2$ and the spin components \emph{without} SOI in Fig.~(\ref{fig:noSOI}).  Since we work with dimensonless parameters, this is the diffraction pattern equivalently at $L=1$.  Furthermore, we consider the wave propagating to the positive $x$ direction.  $|\psi|^2$ we obtained in this case is exactly the same as the diffraction pattern of photon.  Dark fringes locate at the same position as the formula given in all textbooks of general physics.  Without surprise, $|\psi_1|^2$ and $|\psi_1|^2$ are the same.  These results imply that the electron spin remain lies in the $xy$ plane after diffraction.  Actually, without SOI, upper and lower bands are degenerate.  Given a chemical potential, both spin up and spin down pass through the slit.  Therefore, the spin distribution for each component defined by
\begin{eqnarray}
<S^i(y)>=\frac{<\psi(y)|S^i|\psi(y)>}{<\psi(y)|\psi(y)>} \ \ \text{for} \ \ i = x,y,z
\end{eqnarray}
are trivial.  Namely $<S^i(y)>=0$ for all components.

\begin{figure}[htbp] 
   \centering
   \includegraphics[scale=0.2]{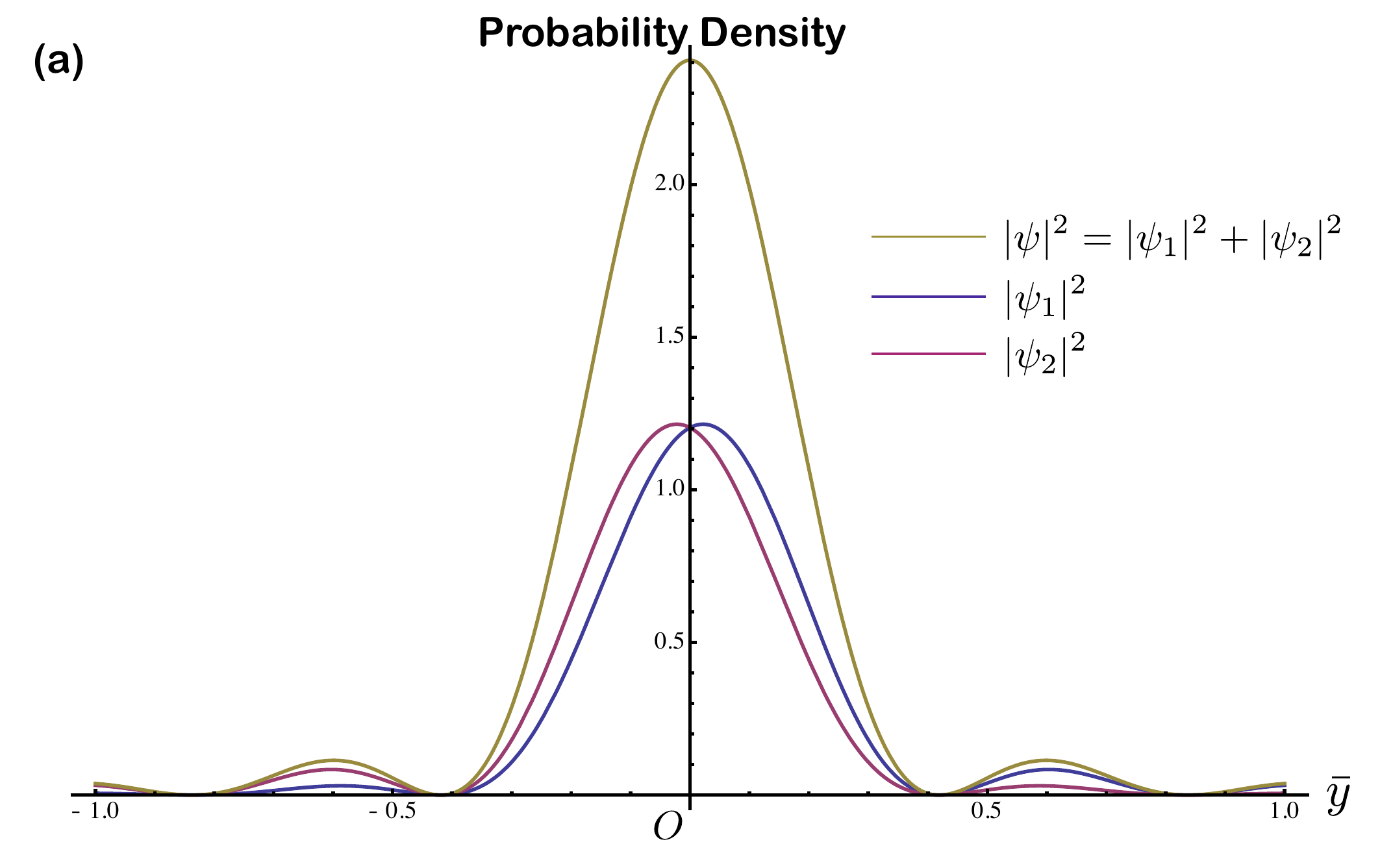}
   \includegraphics[scale=0.2]{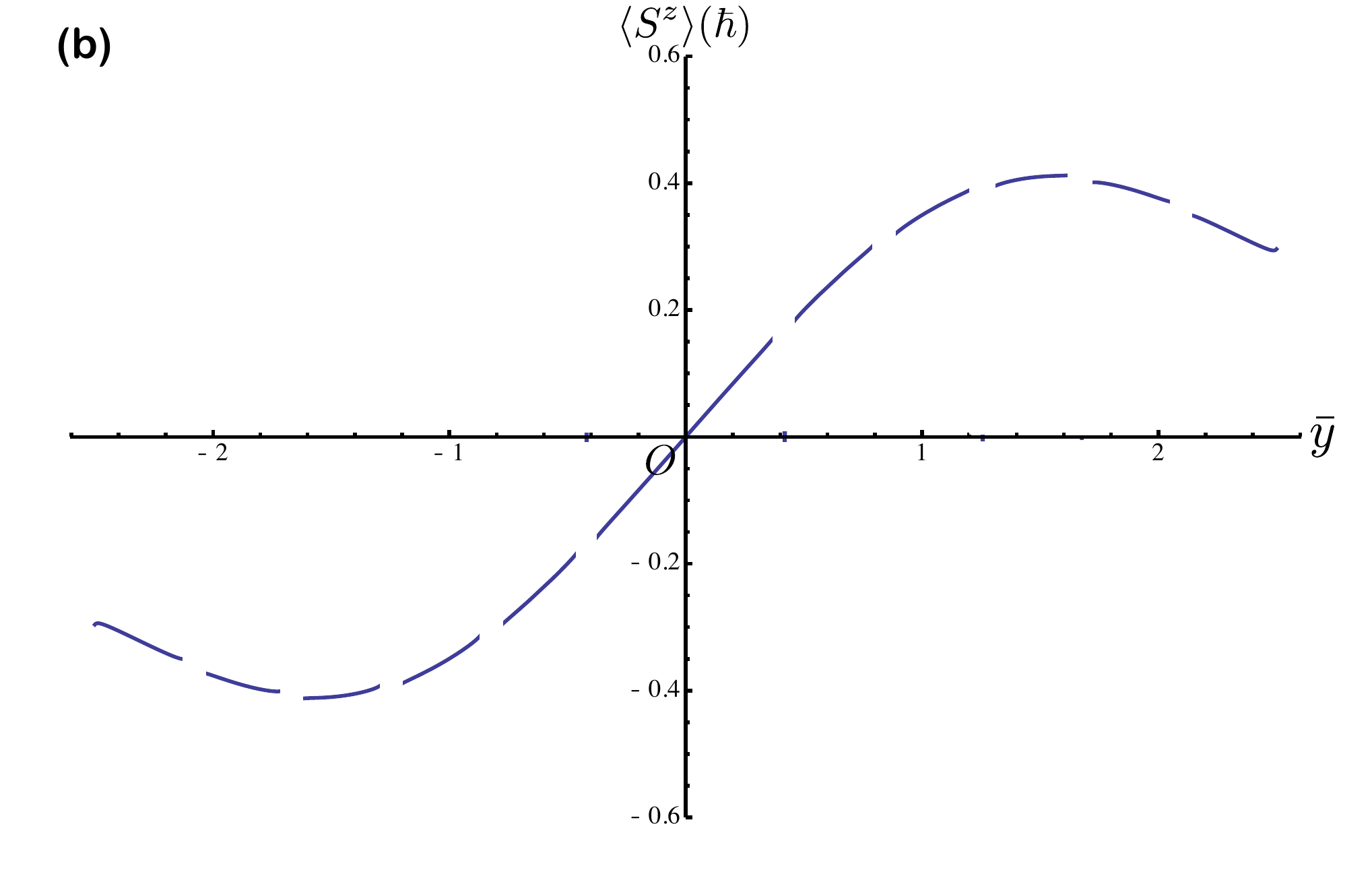}  
   \includegraphics[scale=0.19]{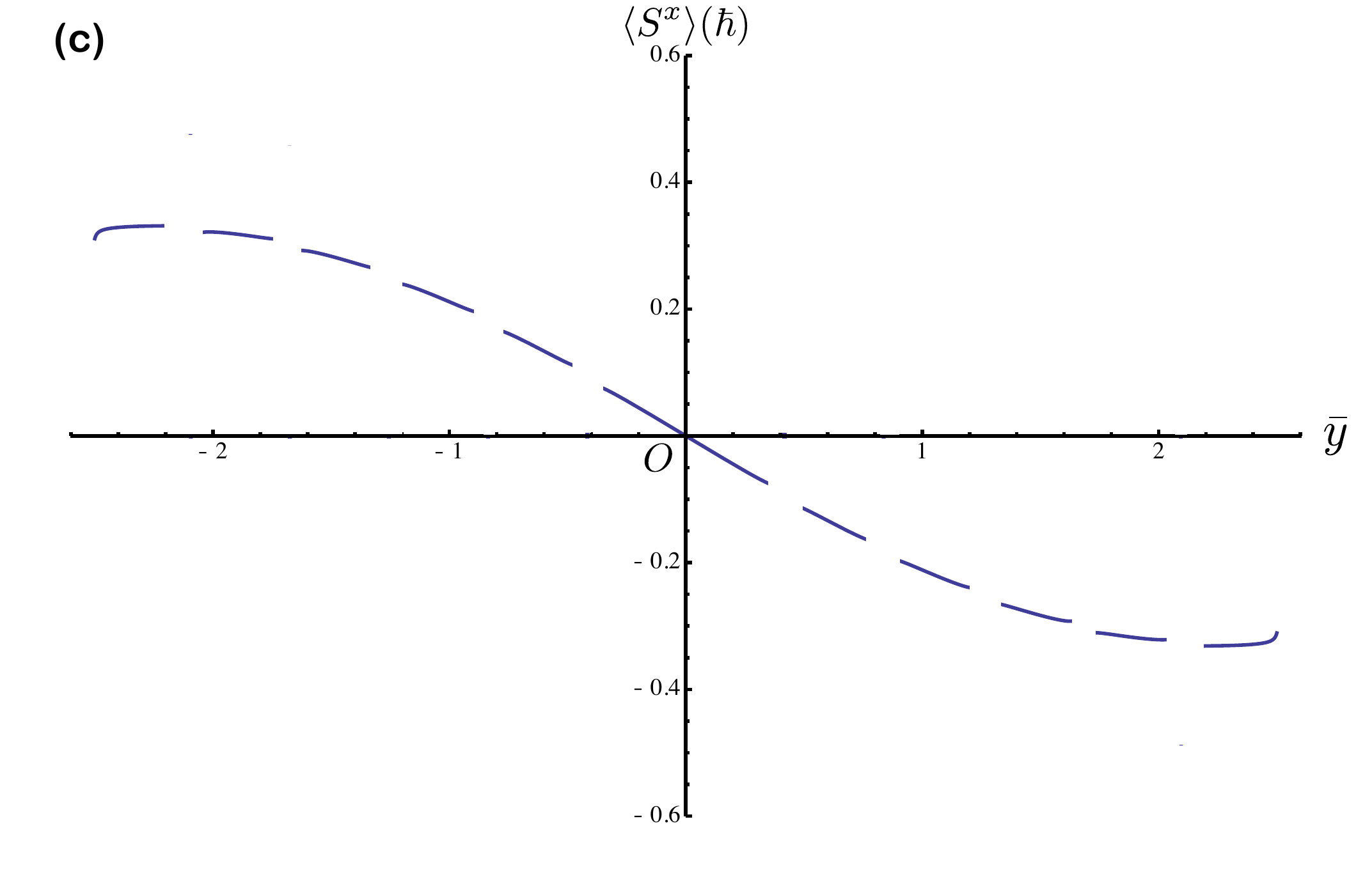}
   \includegraphics[scale=0.2]{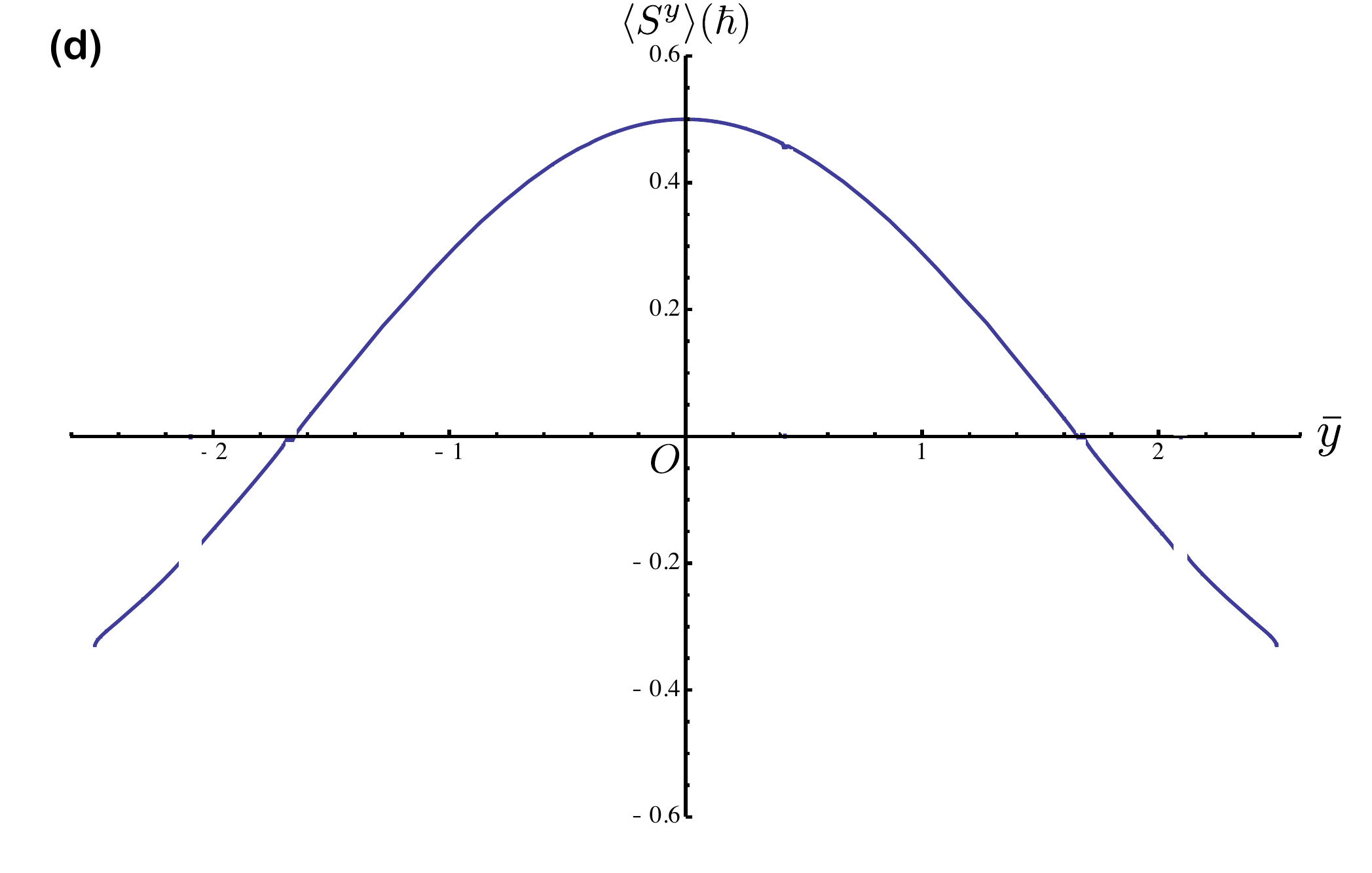} 
   \caption{(color online) The diffraction pattern and spin distribution for the Rashba lower band in the dimensionless parameters $\bar{\alpha}=0.5$, $k=7.5$, and $\bar{d}=0.01$. (a)The yellow line indicates the distribution of probability densities $|\psi|^2$. The blue line is the distribution of $|\psi_1|^2$, and the red line is the one of $|\psi_2|^2$.  (b)(c)(d) are the spin distributions $<S^i(y)>$.}
   \label{fig:rashbad}
\end{figure}

\begin{figure}[htbp] 
   \centering
   \includegraphics[scale=0.2]{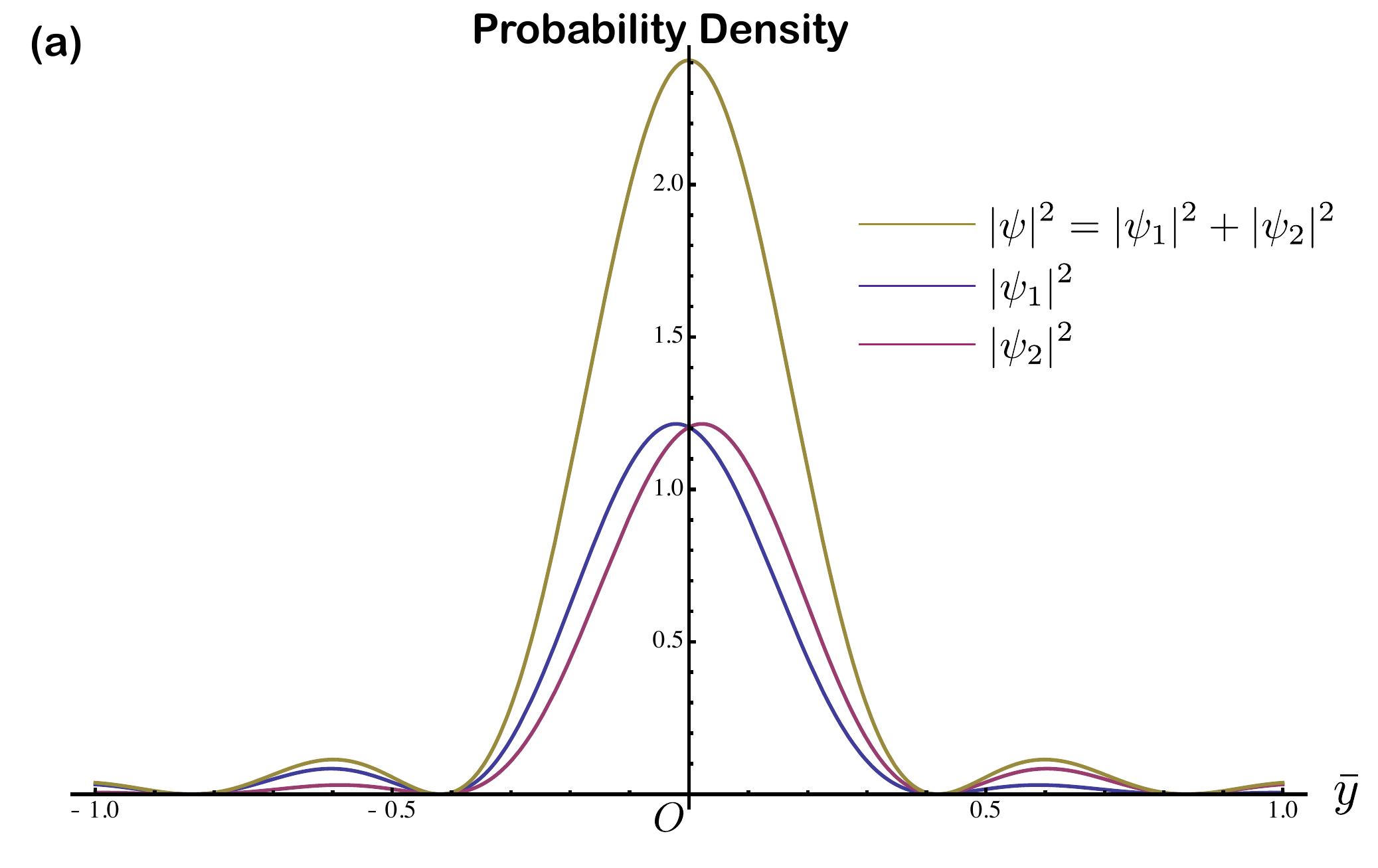} 
   \includegraphics[scale=0.2]{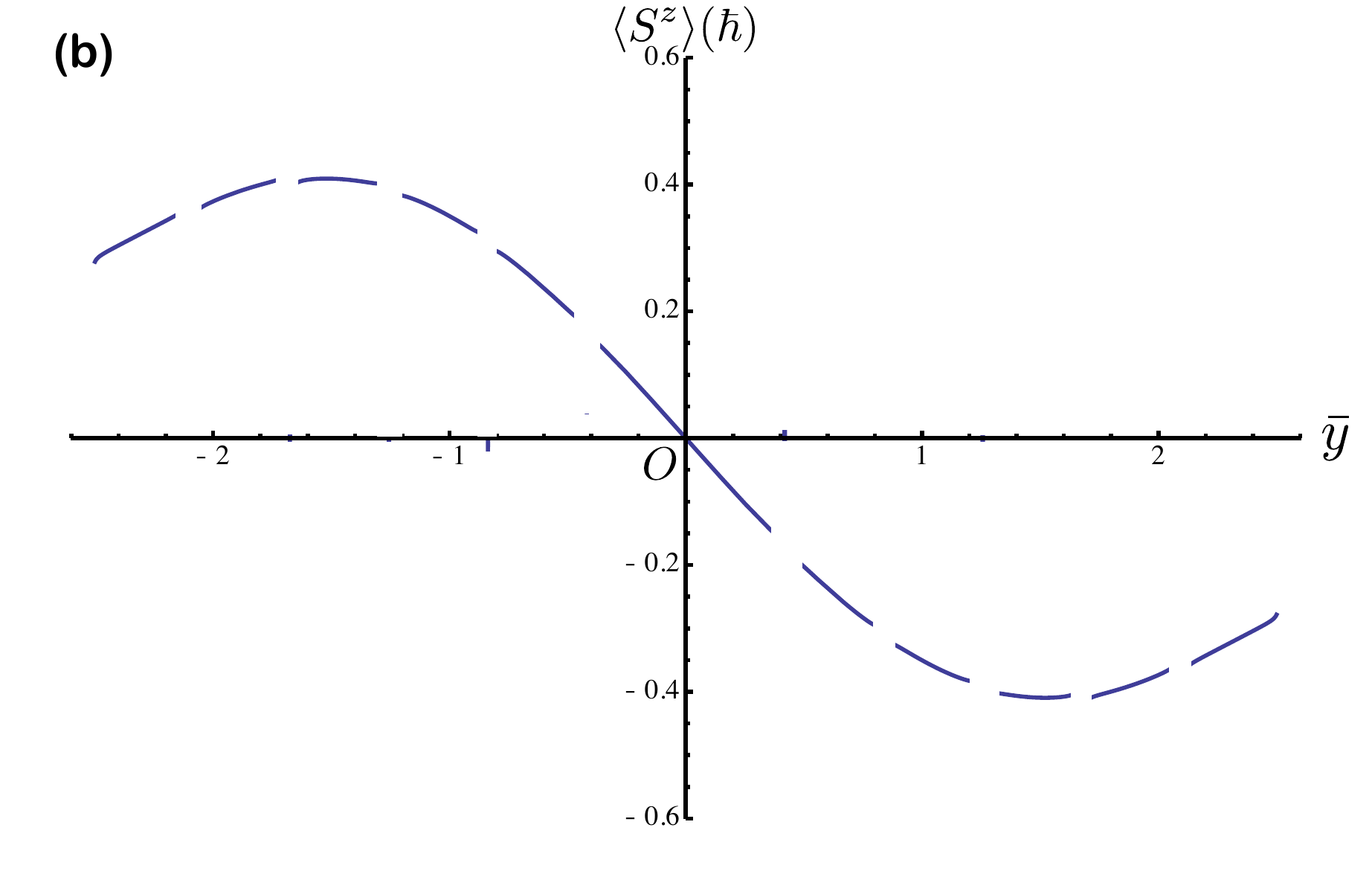}
   \includegraphics[scale=0.2]{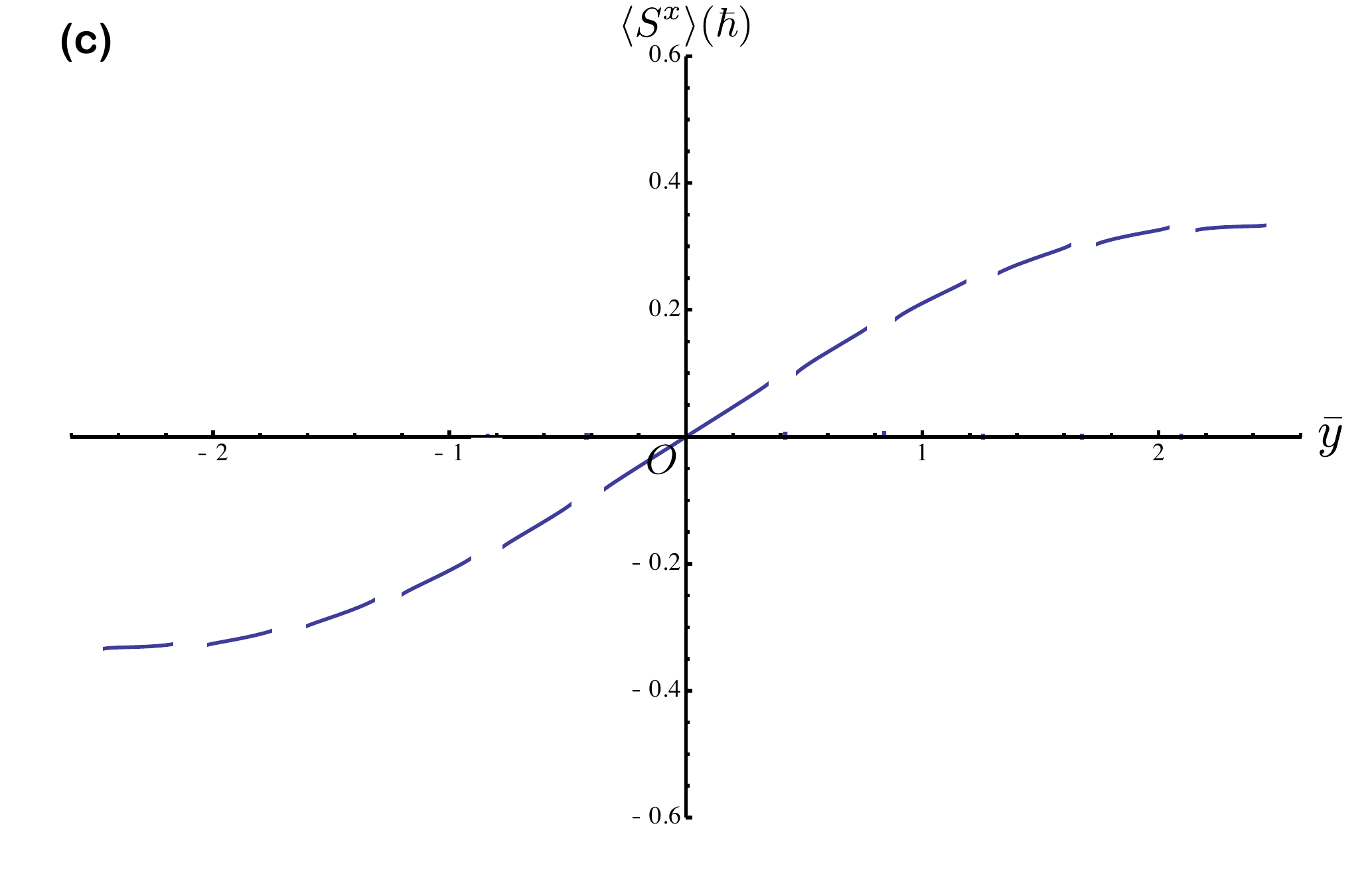}
   \includegraphics[scale=0.2]{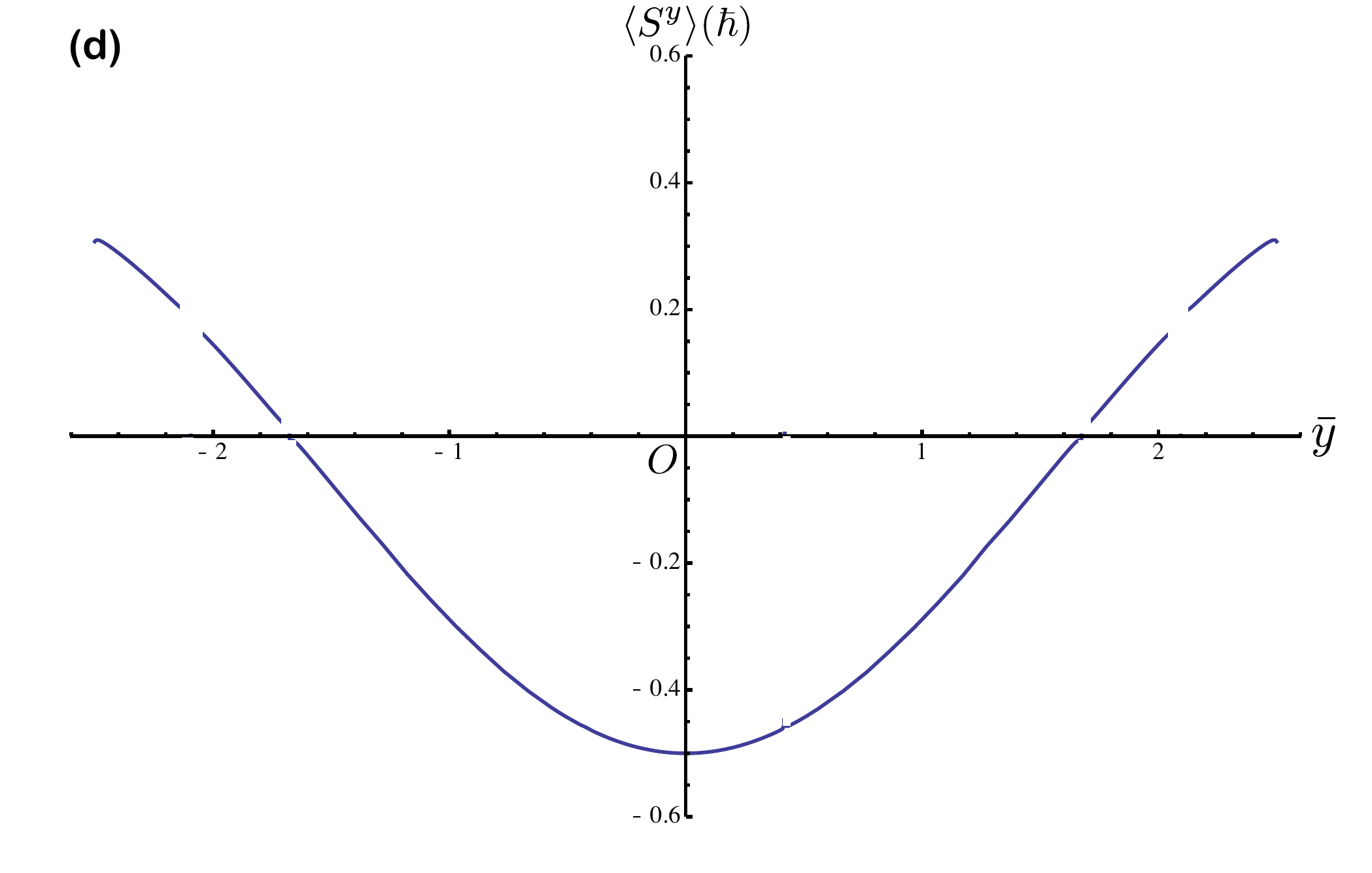}
   \caption{(color online) The diffraction pattern and spin distribution for the Rashba upper band in the dimensionless parameters $\bar{\alpha}=0.5$, $k=7.5$, and $\bar{d}=0.01$. (a)The yellow line indicates the distribution of probability densities $|\psi|^2$. The blue line is the distribution of $|\psi_1|^2$, and the red line is the one of $|\psi_2|^2$.  (b)(c)(d) are the spin distributions $<S^i(y)>$.}
   \label{fig:rashbau}
\end{figure}

When SOI is considered, $<S^i(y)>$ become nontrivial.  In Fig.~(\ref{fig:rashbad}) and Fig.~(\ref{fig:rashbau}), we show the results for the lower and upper bands in the Rashba system respectively.  If the chemical potential is tuned above the band-crossing point, the electrons from the upper band has larger diffraction effect, because they have longer wavelength.  If the chemical potential is tuned below the band-crossing point, the diffraction pattern comes from the electron of the lower band.  Fig.~(\ref{fig:rashbad}a) is the diffraction pattern, namely $|\psi|^2$ drawn in a yellow line, for the lower band.  It is exactly the same as the one without SOI.  If an experimental method that is not spin-resolved, for example scanning probe microscopy, is used, one can not distinguish the difference between the system with or without SOI.  However, the spin components $|\psi_1|^2$ and $|\psi_2|^2$ have asymmetric $y$ dependence.  Their dark fringes locate at the same points.   "The bright fringes" distribute, however, asymetrically.  The asymmetry results in a new effect of spin splitting.  in Fig.~(\ref{fig:rashbad}b) to Fig.~(\ref{fig:rashbad}d), we compute $<S^i(y)>$ for different components.  At $y=0$,  $<S^z(0)>=<S^x(0)>=0$.  The spin orientation on the screen at $y=0$ is the same as the one of the initial wavefunction at the slit.  At $y\ne 0$, the other two components start to develop in a way that is the odd function of $y$.  The spin distribution for the electrons in the upper band is shown in Fig.~(\ref{fig:rashbau}b) to Fig.~(\ref{fig:rashbau}d).  Since the spin components $|\psi_1|^2$ and $|\psi_2|^2$ are asymmetric in the opposite way to the lower band, the spin distribution distribute oppositely.  If we define the spin current as $I^i_j=<S^i>v_i$, our results imply the existence of the spin current in the transverse direction, namely $I^x_y\ne 0$ and $I^z_y\ne 0$.

\begin{figure}[htbp] 
   \centering
   \includegraphics[scale=0.2]{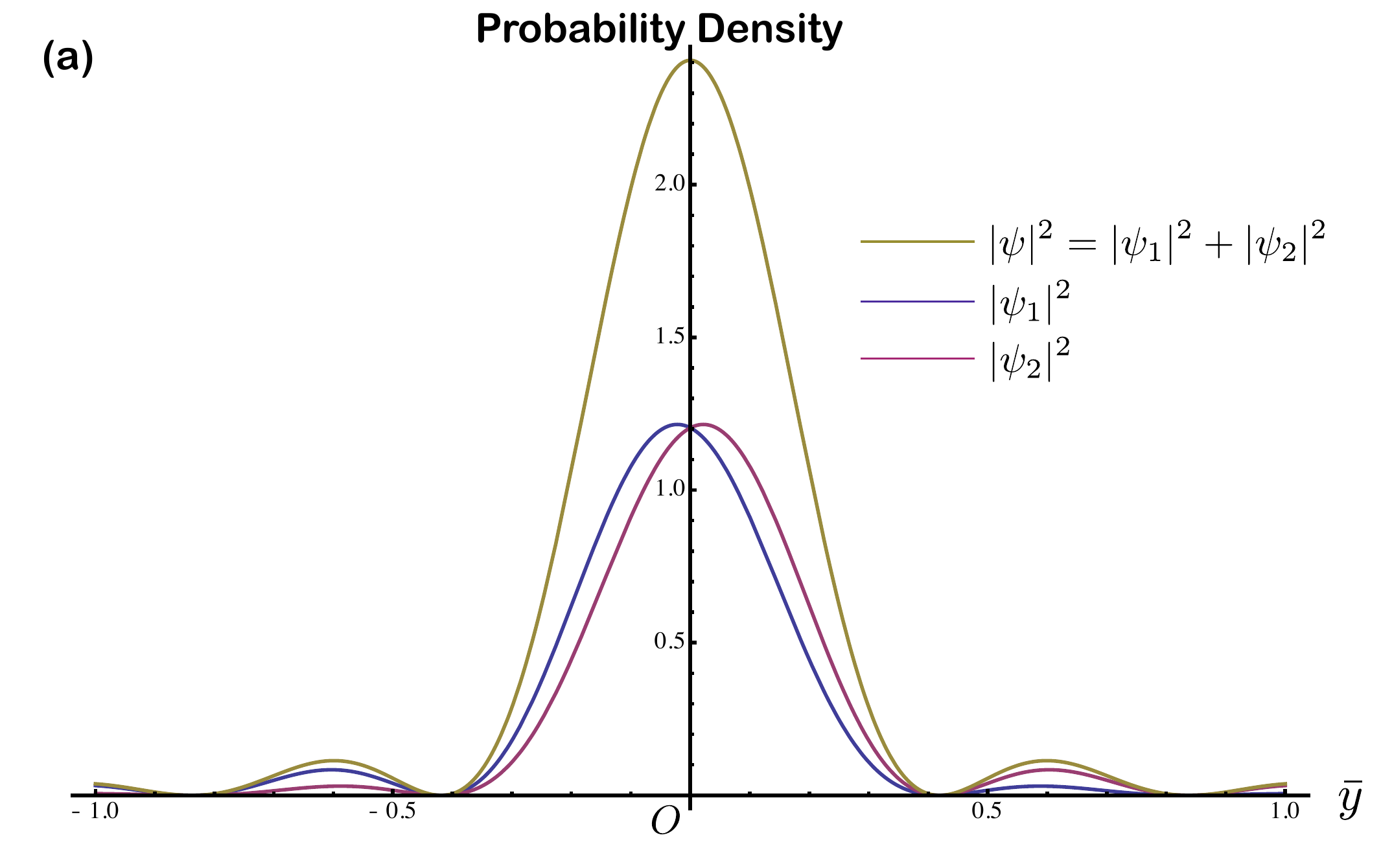} 
   \includegraphics[scale=0.2]{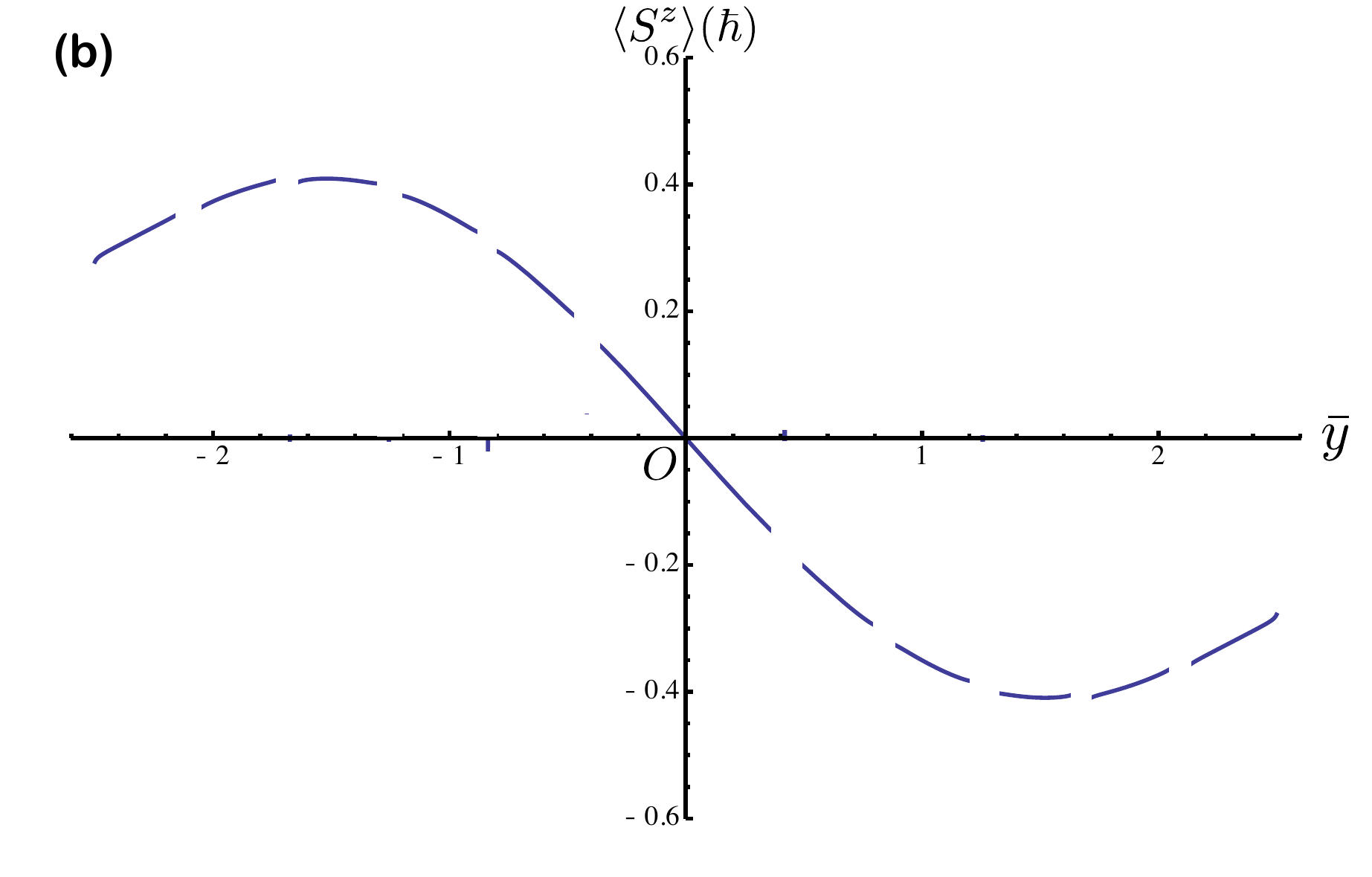}
   \includegraphics[scale=0.2]{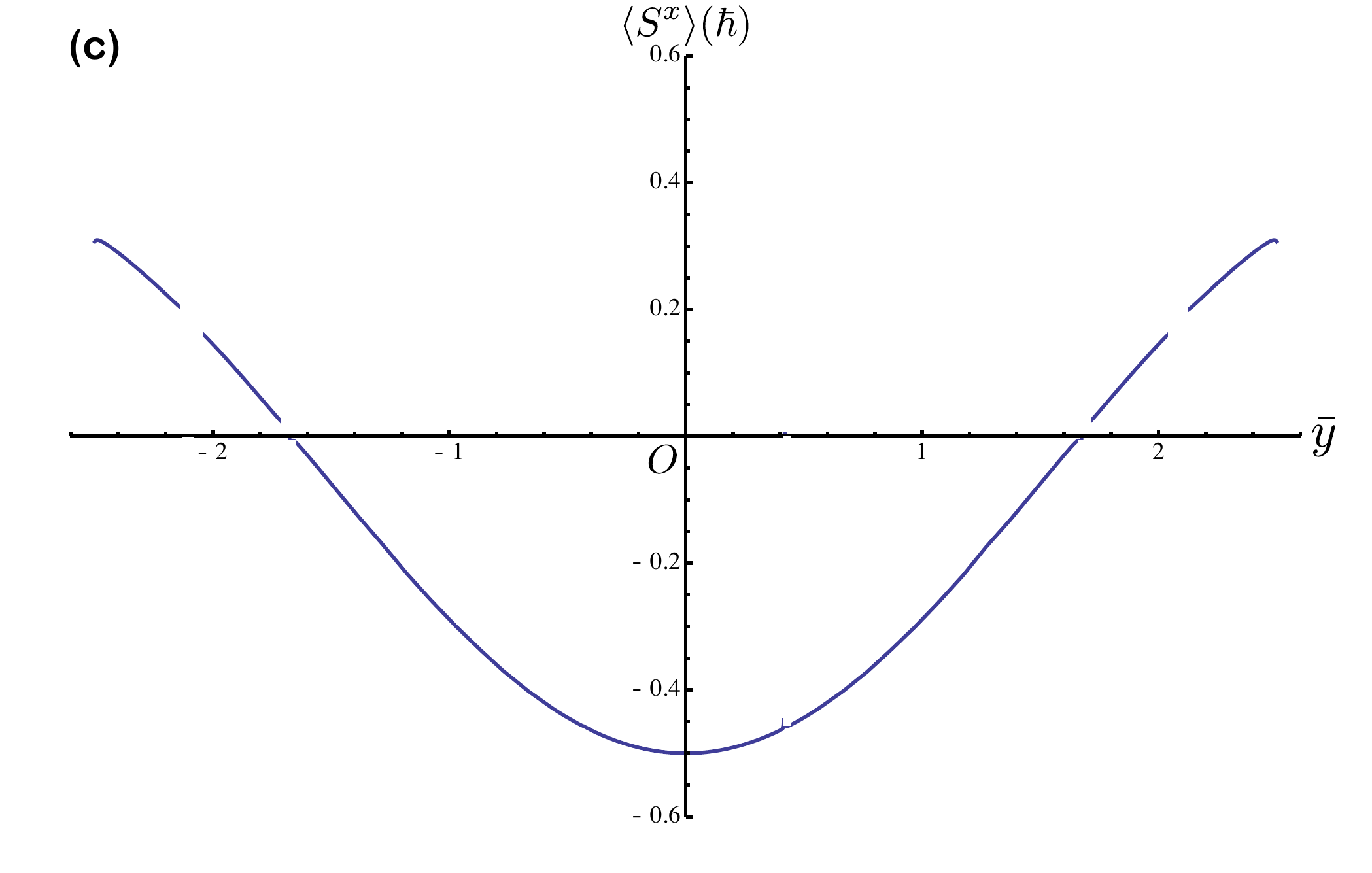}
   \includegraphics[scale=0.2]{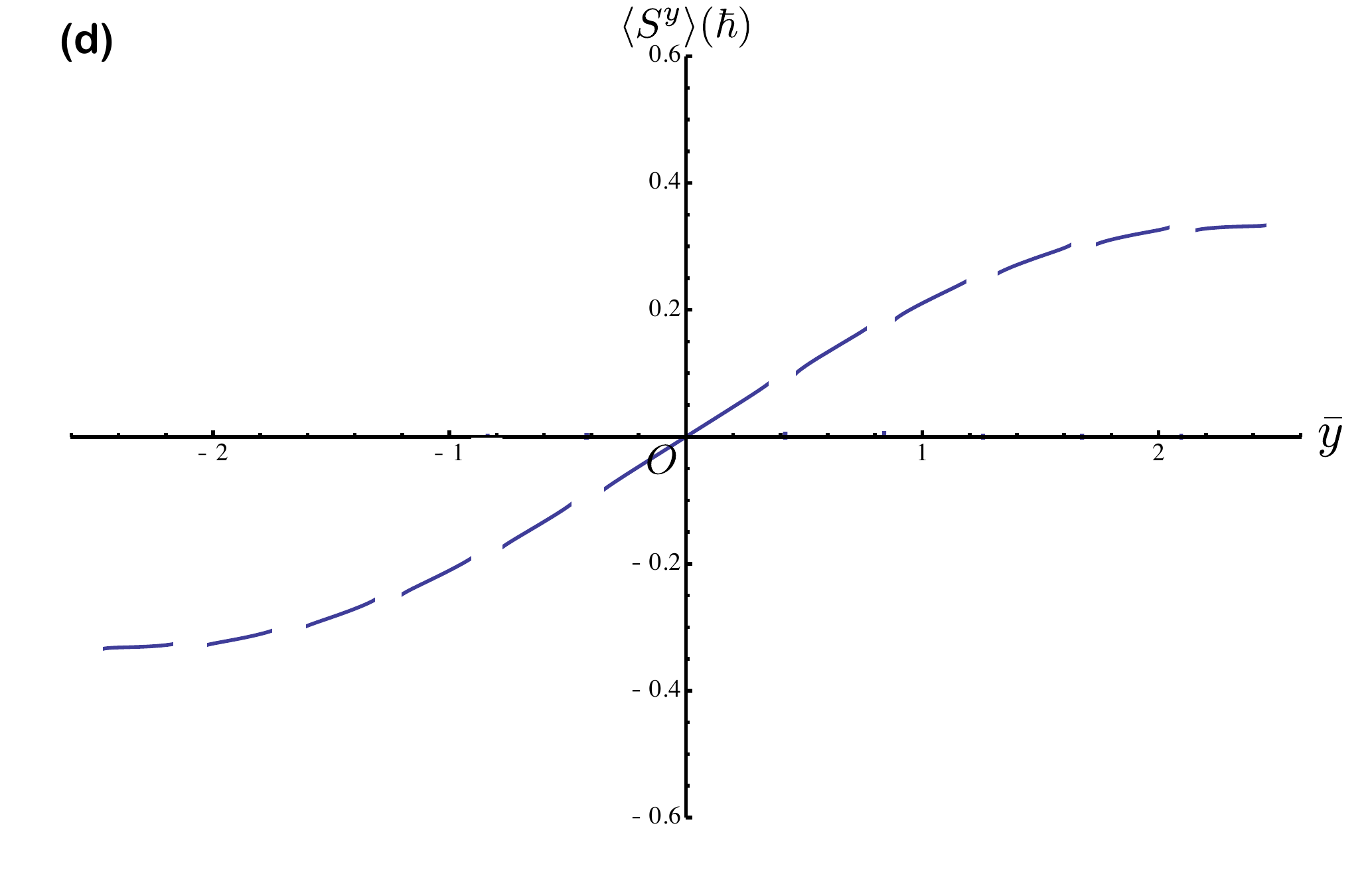}
   \caption{(color online) The diffraction pattern and spin distribution for the Dresselhaus lower band in the dimensionless parameters $\bar{\beta}=0.5$, $k=7.5$, and $\bar{d}=0.01$. (a)The yellow line indicates the distribution of probability densities $|\psi|^2$. The blue line is the distribution of $|\psi_1|^2$, and the red line is the one of $|\psi_2|^2$.  (b)(c)(d) are the spin distributions $<S^i(y)>$.}
   \label{fig:dressd}
\end{figure}

\begin{figure}[htbp] 
   \centering
   \includegraphics[scale=0.2]{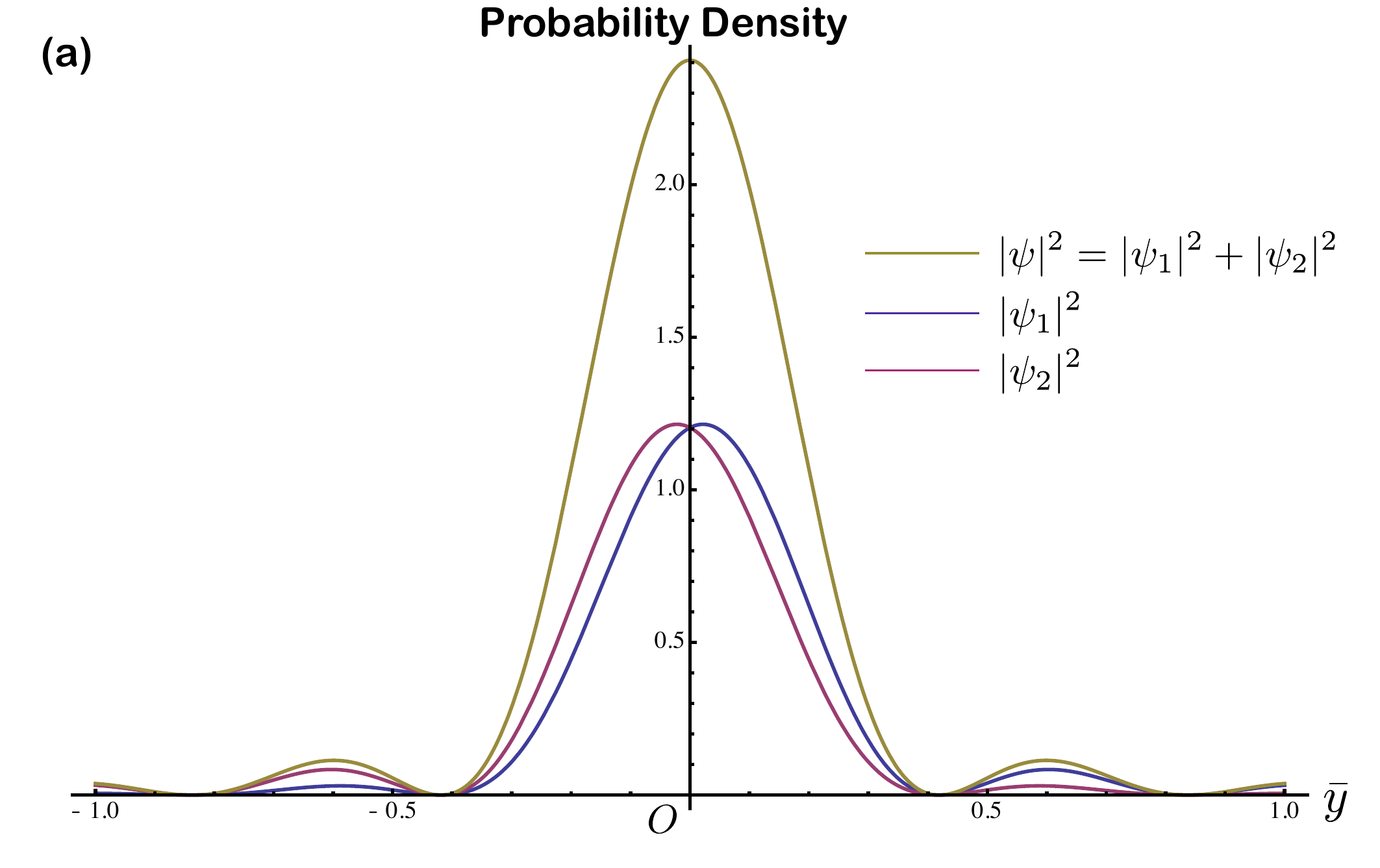} 
   \includegraphics[scale=0.2]{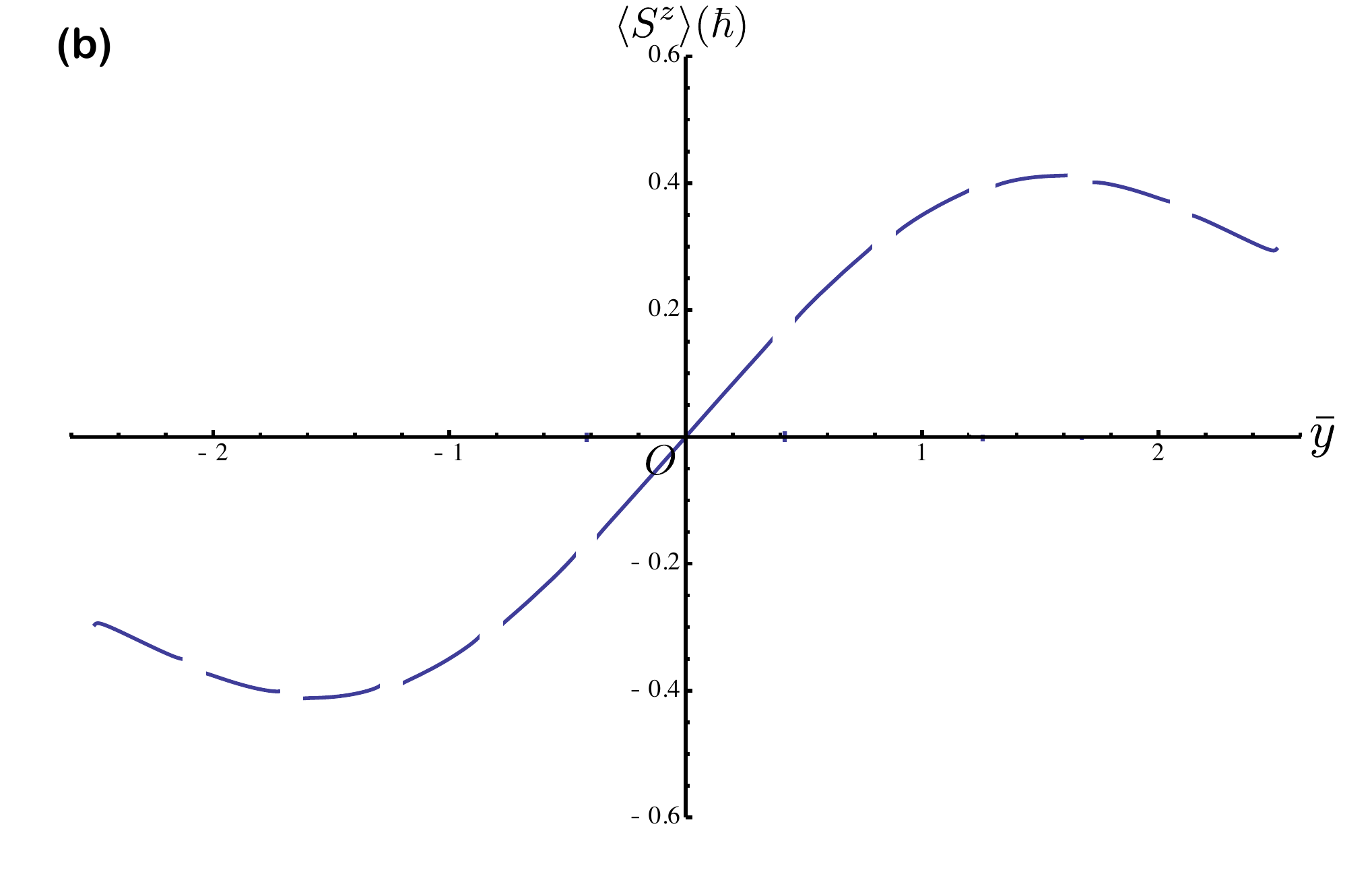} 
   \includegraphics[scale=0.19]{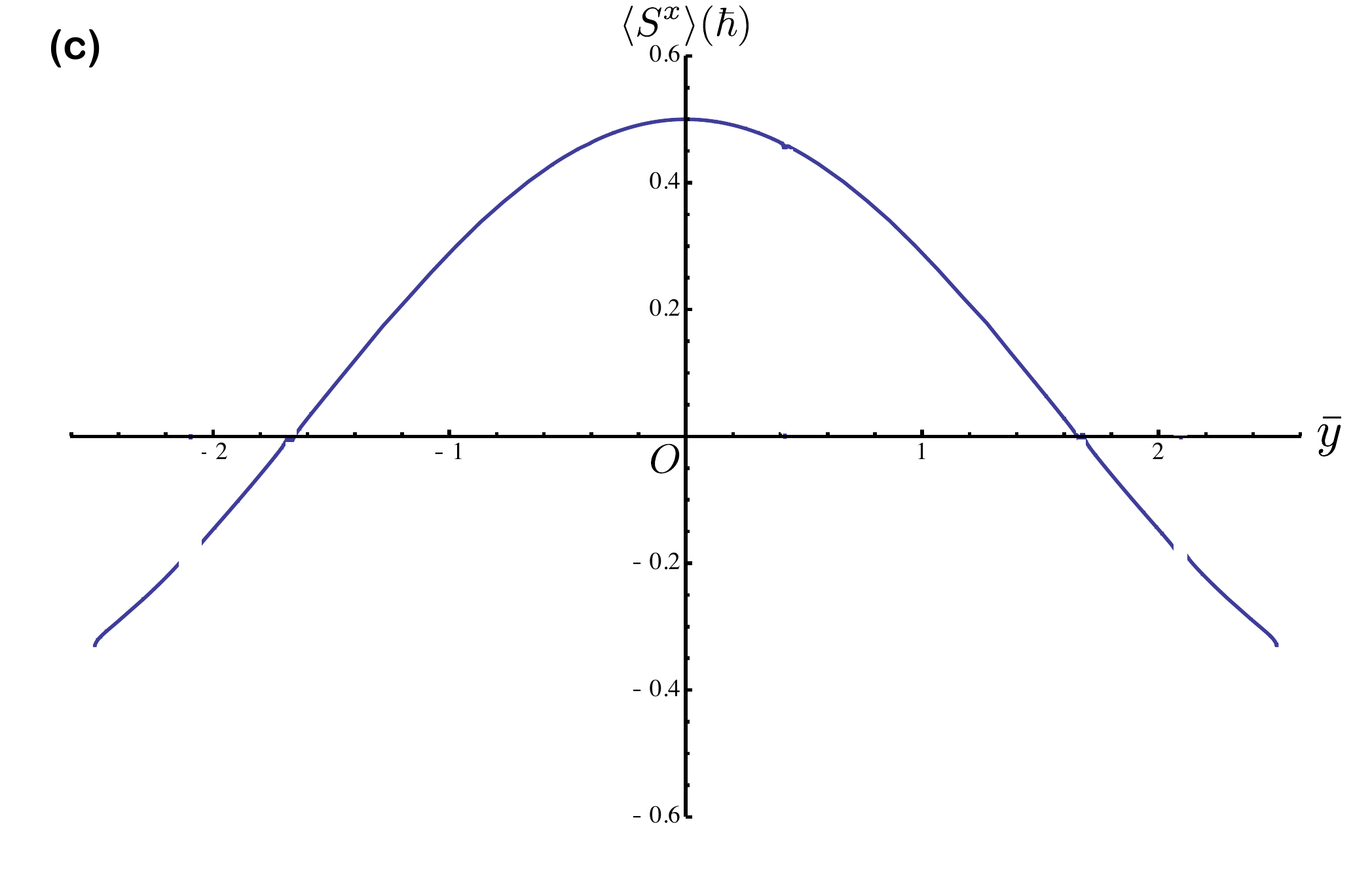} 
   \includegraphics[scale=0.19]{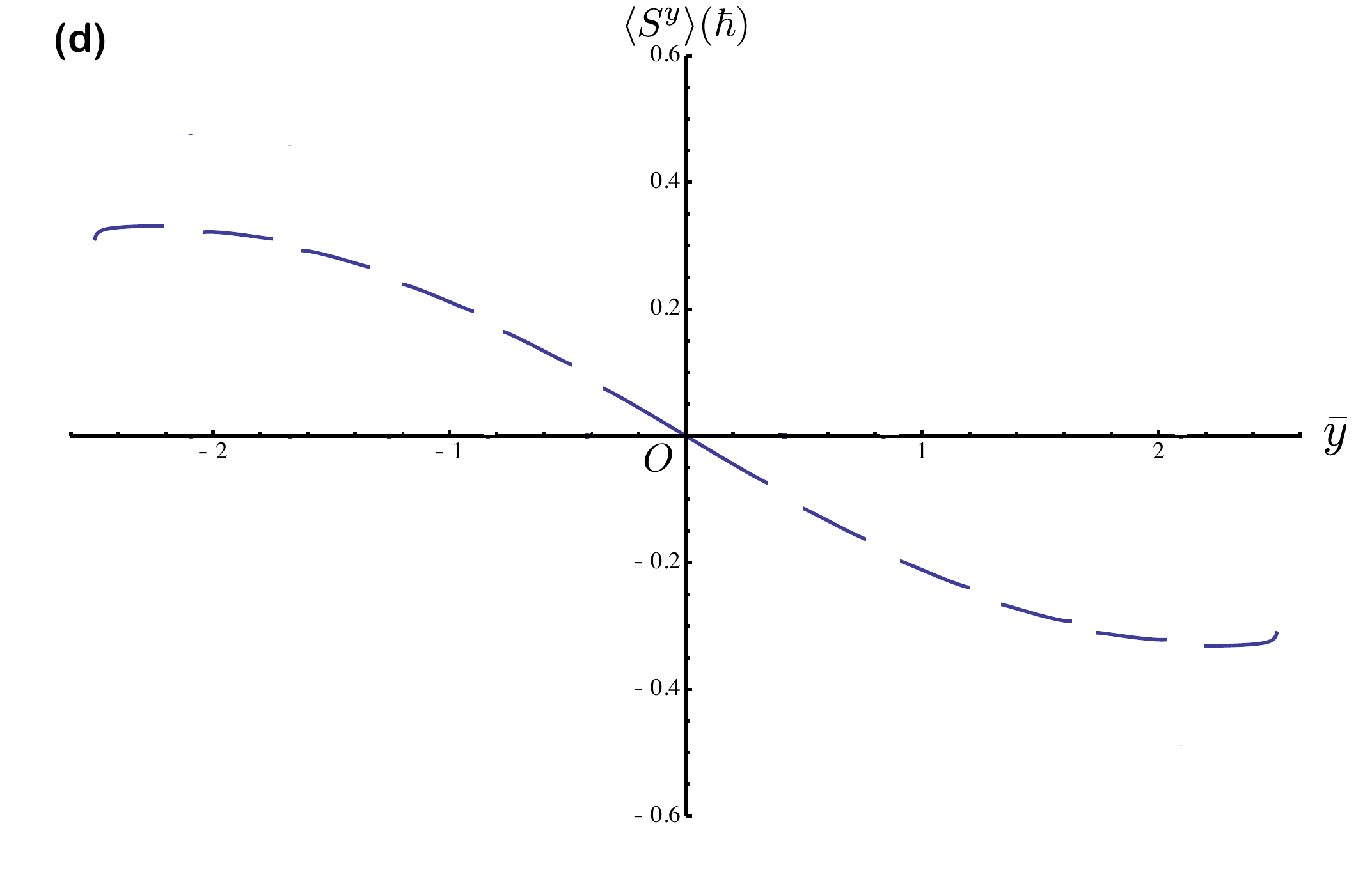} 
   \caption{(color online) The diffraction pattern and spin distribution for the Dresselhaus upper band in the dimensionless parameters $\bar{\beta}=0.5$, $k=7.5$, and $\bar{d}=0.01$. (a)The yellow line indicates the distribution of probability densities $|\psi|^2$. The blue line is the distribution of $|\psi_1|^2$, and the red line is the one of $|\psi_2|^2$.  (b)(c)(d) are the spin distributions $<S^i(y)>$.}
   \label{fig:dressu}
\end{figure}

One important feature in those results is that the $<S^z(y)>$ component becomes finite at $y\ne0$.  One thing we learn in the Rashba system is that the electron spin lies in the $xy$ plane.  After scattered by a slit, a component perpendicular to the $xy$ plane develops!  Its magnitude grows as $|y|$ increases.  Its maximum value can be as large as 0.42 $\hbar$ which is $84\%$ of the $S_z=\hbar/2$.  Another interesting feature in the spin distribution is that there exists $y=y_0$ so that electron spin at $y=y_0$ and $y=-y_0$ are antiparallel.  It is the position where $<S^y(y_0)>=0$, since only $<S^y(y)>$ is an even function in $y$.  We further note that the spin distribution is \emph{independent of} the chemical potential.  It only depends on the SOI strength.  We also compute the spin distribution for the Dresselhaus case.  We provide the numerical results in Fig.~(\ref{fig:dressd}) and Fig.~(\ref{fig:dressu}).

The strength of the signal to detect spin distribution in experiments should be proportional to $<\psi(y)|S^i|\psi(y)>$, which is the product of the spin $<S^i>$ and the wave amplitude.  If the chemical potential is tuned below the band-crossing point, the electrons at the Fermi energy come from the lower band.  In this case, the central maximum of the diffraction peak is spin-polarized.  If the chemical potential is tuned above the band-crossing point, the electron taken part in the diffraction come from both upper and lower bands.  Since the electron spin is polarized in the opposite direction between upper and lower bands.  In the central peak, the signal of spin polarization is almost zero.  In some applications, the spin component perpendicular to the plane is useful.  Although this new spin-splitting effect gives rise to the $S^z$ component, the signal is weak.  In the first diffraction peak, $<S^z>$ reaches its maximum.  However, the wave amplitude is much smaller than the central peak maximum.  Therefore, new idea is needed to enhance the signal of $S^z$.  In the next subsection, we consider the grating device to achieve this goal.

\subsection{Diffraction by grating}\label{dg}

The optical grating is usually made of hundreds or thousands slits that make diffraction peaks more localized.  It also increases the separation distance between the diffraction peaks of the photons of different frequencies.  Therefore, it is one of the basic optical devices to separate photons of different frequencies.  Here, we borrow those beautiful features to design an electronic device to enhance the spin polarization.

\begin{figure}[htbp] 
   \centering
   \includegraphics[scale=0.3]{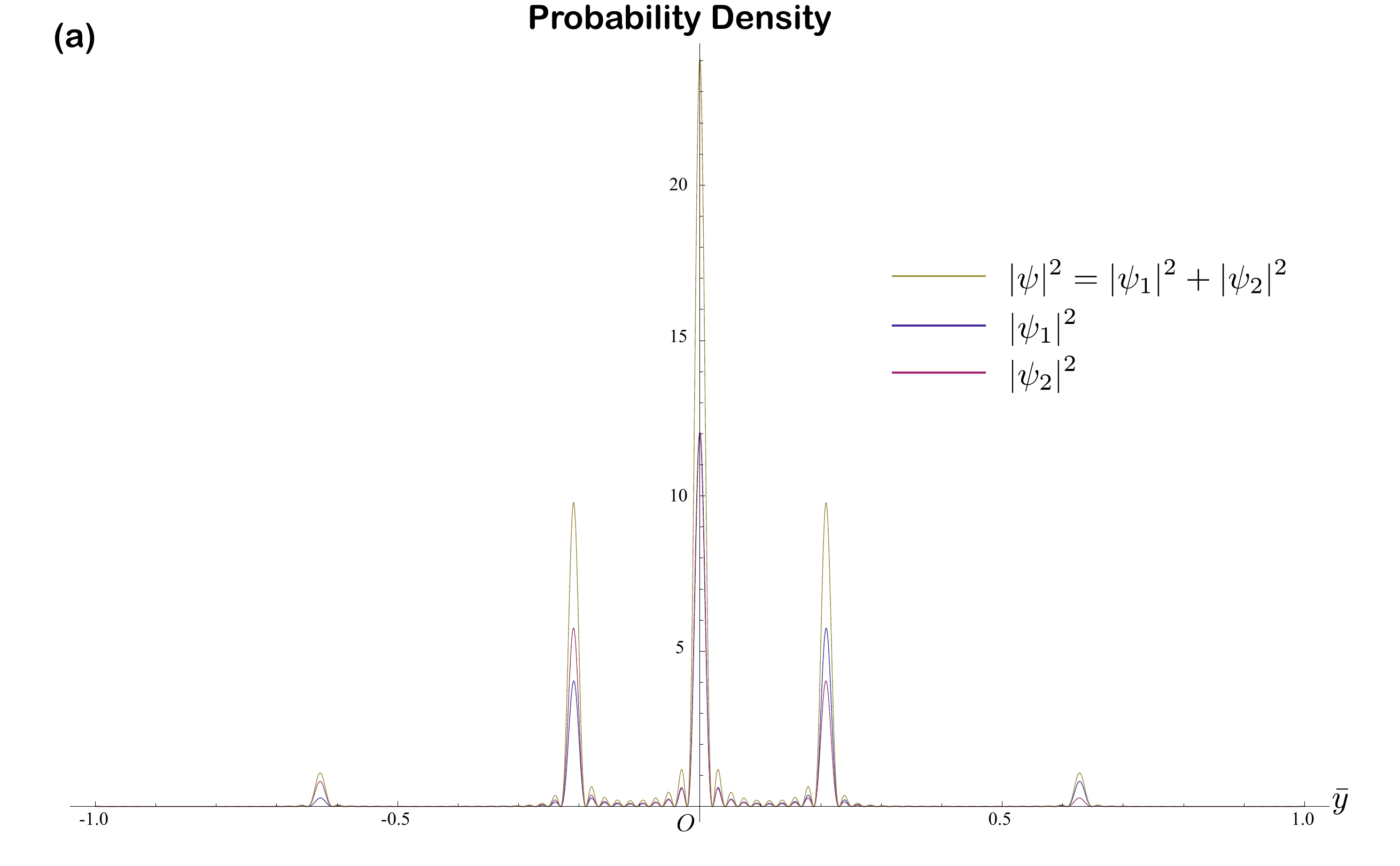}
    \includegraphics[scale=0.3]{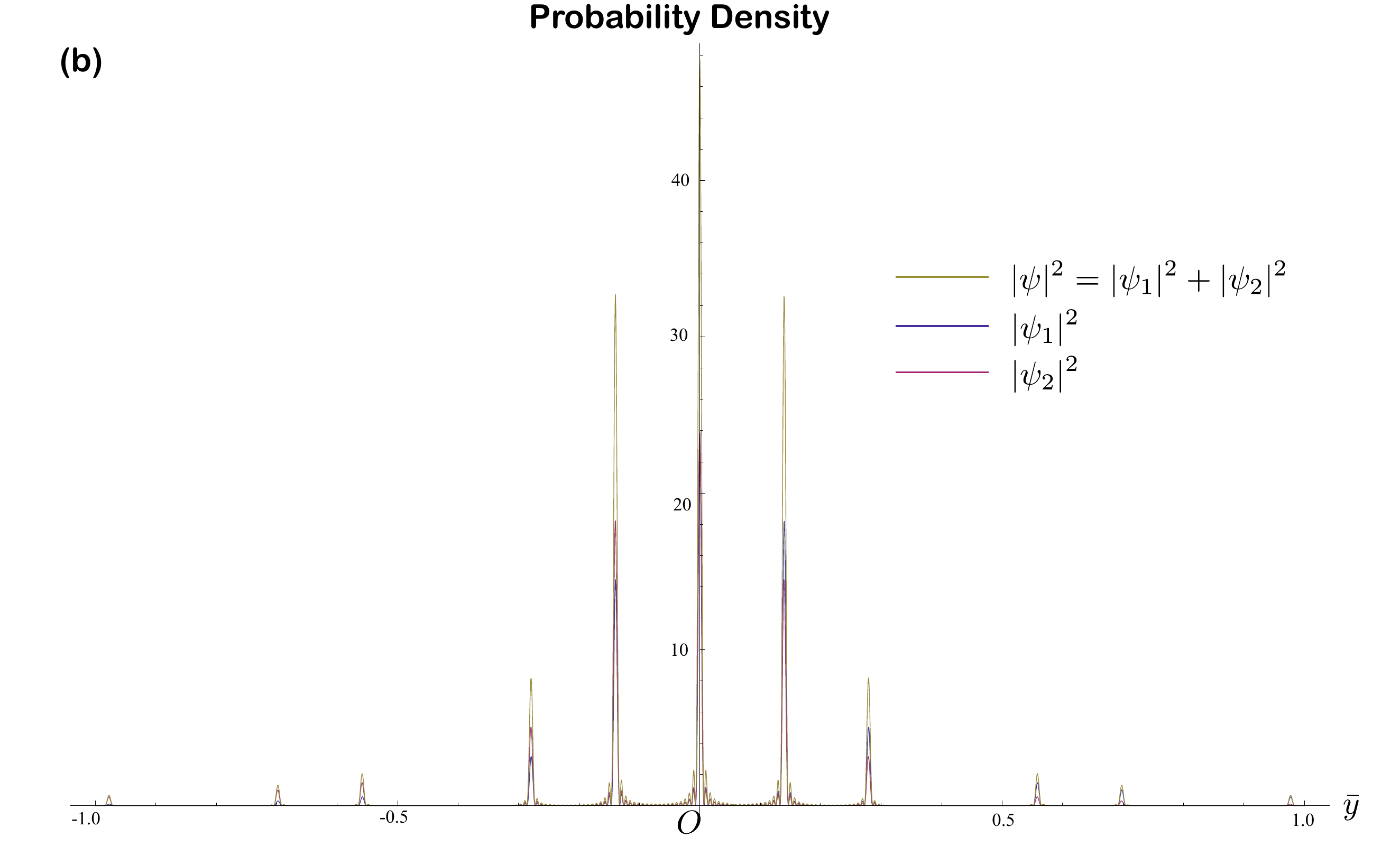}
   \caption{(color online) Distributions of probability density of $|\psi|^2$, $|\psi_1|^2$, and $|\psi_2|^2$for grating devices $N=10$ and 20 in the dimensionless parameters $\bar{\alpha}=0.5, k=7.5$, $\bar{d}=0.01$, and $\bar{b}=0.03$, where $N$ is number of slits, $\bar{b}=b/L$, and $b$ is the separation distance between nearest slits. They are the results of the Rashba lower band.  (a) $N=10$. (b) $N=20$.}
   \label{fig:grating}
\end{figure}

The computational procedure is the same as the one for the single-slit problem.  The only difference is the integration range.  Here, we consider 10 slits and 20 slits with equal distance.  The quantum amplitude on the screen comes from the initial wavefunction at all point sources in all slits.  In Fig.~(\ref{fig:grating}), we show the results of the calculation for $N=10$ and $N=20$ for the lower Rashba band, where $N$ is the number of slits.  The parameters $\bar \alpha (\bar \beta) = 0.5$ corresponding to the real parameter $\alpha \hbar (\beta \hbar)=7.62\times10^{-13} eV\cdot m$ are used and the screen is $L=1\ \mu m$ away.  We reproduce the correct diffraction pattern for grating.  The peak for $N=20$ are much localized and sharper than the one for $N=10$.

The asymmetry of $|\psi_1|^2$ and $|\psi_2|^2$ gives rise to the same spin distribution as the single-slit problem.  It is independent of not only the wavelength of electron but also the number of slits.  We will provide the explanation in the next section.  The position of the diffraction peaks makes no difference from the one of photons.  Different from the case without SOI, the spin orientation of diffraction peaks are different.  One can tune the chemical potential to change the electron wavelength and the position of the diffraction peaks. Since $<S^z(y)>$ is independent of the electron wavelength, one can tune the spin orientation of the diffraction peaks.  In optics, it is the first diffraction peak used to distinguish photons of different frequency.  Here, the first diffraction peak at positive $y$ and negative $y$ are different.  At $y=y_0$ mentioned in the single-slit case, the spin orientation of two first diffraction peaks is \emph{antiparallel}.  When the chemical potential is tuned so that the first diffraction peaks locates at $y=|y_0|$,  
one can observe opposite spin orientations on the screen, mimicking the Stern-Gerlach experiment.

\section{Discussion and conclusion}\label{discussion}

As mentioned in the introduction, the spin-splitting effect proposed in the paper is due to the precession of electron spin subjected to the fictitious magnetic field by the SOI.  After scattered by the slit, electron acquires a transverse momentum, namely the $y$ direction.  The electrons propagating in the positive $y$ and negative $y$ precess in opposite directions, and hence the spin distribution arises.  Due to the precession mechanism, the spin distribution depends on the distance of the screen from the slit.  However, it does not depend on the electron wavelength.  The reason is giving as the following.  Changing the wavelength means to change the electron momentum.  For example, if the electron has larger momentum, it takes shorter time to reach the screen.  On the other hand, the magnitude of the fictitious magnetic field is proportional to the electron momentum.  The larger the momentum is, the larger the fictitious magnetic field.  Even though the electron with larger momentum reaches the screen faster, it also has larger rate of precession.  Therefore, the final spin orientation becomes independent of the electron wavelength.  

The precession mechanism also provides the reason why the spin distribution does not depend on the number of slits, namely gratings.  It depends only on the strength of SOI and where to measure it, namely distance of screen.  Within spin coherent length, one can use this new effect to manipulate electron spin.  Usually, the coherent transport can be observed more easily in the low temperature.  In our system, the spin coherent transport can be achieved more easily if there is no interband mixing, that mixes electrons of opposite spins.  Interband mixing can happen by the thermal excitation.  In usual cases, energy to cause interband mixing is around few $me$V, which corresponds to few tens of kelvin.

A good feature of current proposal is that the diffraction pattern is exactly the same as the one of photon even when the SOI is present.  The diffraction nature depend hundred percents on the wave nature of electron.  The new thing we discovered is the non-trivial spin distribution in the diffraction pattern due to the \emph{spinor} nature of electrons and the SOI.  It brings huge convenience to control the peak of the diffraction peaks without worrying the strength of SOI.  The experimental feasibility is already discussed in our precious Letter and will not repeat here.

In conclusion, we propose a new effect of spin-splitting in the 2DEG in the presence of SOI.  It is based on the electron spin precession due to the fictitious magnetic field generated by the SOI.  Diffracted by slits, electron gains transverse momentum and processes in opposite ways in the different transverse momentum.  Spin current $I^x_y$ and $I^z_y$ is generated for the Rashba case and $I^y_y$ and $I^z_y$ is generated for the Dresselhaus case.  The presence of the spin current leads to nontrivial spin distribution in the diffraction pattern.  Most importantly, the new effect shed lights on the potential applications in spintronics.

We appreciate the stimulating discussion with Chi-Te Liang for various experimental aspects.  This work was supported by the National Science Council of Taiwan under NSC 97-2112-M-002-027-MY3.

\appendix
\section{Spin distribution in the Dresselhaus case}\label{section1}
The diagonalization of Eq. (\ref{hami}) with $\alpha=0$ and $\beta \neq 0$ gives $H=\Lambda H_{diag}\Lambda^{\dag}$, where
\begin{align}
H_{diag}=&\left(
\begin{array}{cc}
(\frac{p^2}{2m^*}-\mu-\beta p)  &   0  \\
 0 &   (\frac{p^2}{2m^*}-\mu+\beta p)   \\  
\end{array}
\right),\\
\Lambda=&\frac{1}{\sqrt{2}}\left(\begin{array}{cc}   e^{i\theta}& e^{i\theta}\\-1&1\end{array}\right), \ \theta=\tan^{-1}({p_y}/{p_x}).
\end{align}

The propagator in the momentum space is given by
\begin{eqnarray}
&\!\!\!&U(t)=e^{-\frac{i}{\hbar}Ht}=\Lambda e^{-\frac{i}{\hbar}H_{diag}t}\Lambda^\dagger \nonumber\\
&\!\!\!&=\left(\!\begin{array}{cc}\cos(\frac{\beta p}{\hbar}t)&-ie^{i\theta}\sin(\frac{\beta p}{\hbar}t)\\-ie^{-i\theta}\sin(\frac{\beta p}{\hbar}t)&\cos(\frac{\beta p}{\hbar}t)\end{array}\!\right)e^{-\frac{i}{\hbar}(\frac{p^2}{2m}-\mu)t}\nonumber\\&\!\!\!&=\left(\begin{array}{cc}U_{11}&U_{12}\\U_{21}&U_{22}\end{array}\right).
\end{eqnarray}
The kernels in the Dresselhaus case can be written in terms of hypergeometric function given by
\begin{eqnarray}
&\!\!\!&\langle x,y,t|x',y', 0\rangle_{11}=\langle x,y,t|x',y', 0\rangle_{22}\nonumber\\
&\!\!\!&=\left(\frac{m}{2\pi it\hbar}\right)e^{\frac{i}{\hbar}\mu t}\sum_{n=0}^{\infty}\frac{n!}{(2n)!}\left(\frac{2im\beta^2t}{\hbar}\right)^n\nonumber\\
&\!\!\!&\times\ _1F_1(n+1;1;\frac{im\tilde r^2}{2\hbar t}),
\end{eqnarray}
\begin{align}
\langle x,y,t|x',y', 0\rangle_{12}
=&\left(\frac{m}{2\pi it\hbar}\right)\left(\frac{m\beta}{\hbar}\right)e^{\frac{i}{\hbar}\mu t}(-i\tilde x+\tilde y)\notag\\
&\times\sum_{n=0}^{\infty}\frac{(n+1)!}{(2n+1)!}\left(\frac{2im\beta^2t}{\hbar}\right)^n\notag\\&\ _1F_1(n+2;2;\frac{im\tilde r^2}{2\hbar t}),
\end{align}and
\begin{align}
\langle x,y,t|x',y', 0\rangle_{21}=&-\left(\frac{m}{2\pi it\hbar}\right)\left(\frac{m\alpha}{\hbar}\right)e^{\frac{i}{\hbar}\mu t}(i\tilde x+\tilde y)\notag\\
&\times\sum_{n=0}^{\infty}\frac{(n+1)!}{(2n+1)!}\left(\frac{2im\alpha^2t}{\hbar}\right)^n\notag\\& _1F_1(n+2;2;\frac{im\tilde r^2}{2\hbar t}).
\end{align}

For the electron comes from the lower band, the initial wave function is $\phi_-(0,y',0)=(1,-1)^T/\sqrt{2d}$. We compute the matrix product in Eq.~(\ref{wavef}) as the following
\begin{align}
&\int^{d/2}_{-d/2}\langle x,y,t|x',y', 0\rangle_{11}\phi_1(0,y',0)\notag\\
&=\frac{1}{\sqrt{2d}}\left(\frac{m}{2\pi it\hbar}\right)e^{\frac{i}{\hbar}\mu t}
\sum_{n=0}^{\infty}\frac{n!}{(2n)!}\left(\frac{i\bar \beta^2\bar d}{k}\right)^n\notag\\
&\times\frac{1}{2\pi i}\oint\mathrm ds(s-1)^{-(n+1)}s^{n}\int_{-d/2}^{d/2}\mathrm dy'\exp(s\frac{im\tilde r^2}{2\hbar t})  \notag\\
&=\frac{1}{\sqrt{2d}}\left(\frac{md}{2\pi it\hbar}\right)e^{\frac{i}{\hbar}\mu t}F(y),\end{align}
where we used an approximation $\tilde r^2 =x^2+y^2-2yy'+y'^2\approx r^2-2yy'$, $r^2=x^2+y^2$ in the $y'$ integration and the functions
\begin{align}
F(y)=&\sum_{n=0}^{\infty}\frac{1}{(2n)!}\left(\frac{i\bar \beta^2\bar d}{k}\right)^nf_n(y),\\
f_n(y)=&\left[\left(\frac{d}{ds}\right)^ns^{n-1}\exp(s\frac{imr^2}{2\hbar t})\frac{\sin(sk\bar y)}{k \bar y}\right]_{s=1}.
\end{align}
Furthermore,
\begin{align}
&\int^{d/2}_{-d/2}\!\langle L,y,t|x',y', 0\rangle_{12}\phi_2(0,y',0)\!=\!\frac{-1}{\sqrt{2d}}\left(\frac{m}{2\pi it\hbar}\right)e^{\frac{i}{\hbar}\mu t}\notag\\
&\times\left(\frac{m\beta}{\hbar}\right)\sum_{n=0}^{\infty}\frac{(n+1)!}{(2n+1)!}\left(\frac{i\bar \beta^2\bar d}{k}\right)^n\frac{\Gamma(n+1)}{\Gamma(n+2)}\notag\\
&\times\frac{1}{2\pi i}\oint\mathrm ds(s-1)^{-(n+1)}s^{n+1}\notag\\
&\times\int_{-d/2}^{d/2}\mathrm dy'(-iL+y-y')\exp(s\frac{im\tilde r^2}{2\hbar t})  \notag\\
&=\frac{1}{\sqrt{2d}}\left(\frac{md}{2\pi it\hbar}\right)e^{\frac{i}{\hbar}\mu t}\bar \beta\left[(i-\bar y)G(y)+i\bar dH(y)\right],
\end{align}
where
\begin{eqnarray}
&\!\!\!&G(y)=\sum_{n=0}^{\infty}\frac{1}{(2n+1)!}\left(\frac{i\bar \beta^2\bar d}{k}\right)^ng_n(y),\nonumber\\
&\!\!\!&g_n(y)=\left[\left(\frac{d}{ds}\right)^ns^{n}\exp(s\frac{imr^2}{2\hbar t})\frac{\sin(sk\bar y)}{k \bar y}\right]_{s=1},\nonumber\\
&\!\!\!&H(y)=\sum_{n=0}^{\infty}\frac{1}{(2n+1)!}\left(\frac{i\bar \beta^2\bar d}{k}\right)^nh_n(y),\nonumber\\
&\!\!\!&h_n(y)\!=\!\!\left[\!\left(\frac{d}{ds}\right)^n\!\!\!\!s^{n-1}e^{s\frac{imr^2}{2\hbar t}}\!\frac{sk\bar y\cos(sk\bar y)\!\!-\!\sin(sk\bar y)}{2k^2 \bar y^2}\right]_{\!s=1}.
\end{eqnarray}
For $\psi_2$, we have
\begin{eqnarray}
&\!\!\!\!&\int^{d/2}_{-d/2}\!\!\!\langle x,y,t|x'\!,y'\!, 0\rangle_{22}\phi_2(0,y'\!,\!0)\!\!=\!\!\frac{-1}{\sqrt{2d}}\!\left(\!\frac{md}{2\pi it\hbar}\!\right)\!\!e^{\frac{i}{\hbar}\mu t}F(y),\nonumber\\&&
\end{eqnarray}
and
\begin{align}
&\int^{d/2}_{-d/2}\!\!\langle L,y,t|x',y', 0\rangle_{21}\phi_1(0,y',0)\!=\!\frac{-1}{\sqrt{2d}}\left(\frac{m}{2\pi it\hbar}\right)e^{\frac{i}{\hbar}\mu t}\notag\\
&\times\left(\frac{m\beta}{\hbar}\right)\sum_{n=0}^{\infty}\frac{(n+1)!}{(2n+1)!}\left(\frac{i\bar \beta^2\bar d}{k}\right)^n\frac{\Gamma(n+1)}{\Gamma(n+2)}\notag\\
&\times\frac{1}{2\pi i}\oint\mathrm ds(s-1)^{-(n+1)}s^{n+1}\notag\\
&\times\int_{-d/2}^{d/2}\mathrm dy'(iL+y-y')\exp(s\frac{im\tilde r^2}{2\hbar t})  \notag\\
&=\frac{-1}{\sqrt{2d}}\left(\frac{md}{2\pi it\hbar}\right)e^{\frac{i}{\hbar}\mu t}\bar \beta\left[(i+\bar y)G(y)-i\bar dH(y)\right],
\end{align}
where $F(y)$, $G(y)$, and$H(y)$ are given by Eq. (34), Eq. (36), and Eq.(38), respectively.
Therefore,, the electron wave function on the screen $\psi^{(-)}=(\psi_1^{(-)},\psi_2^{(-)})$ in the Dresselhaus lower band can be simplified as the following
\begin{align}
\psi_1^{(-)}=&A\left[F(\bar y)+\bar \beta(i-\bar y)G(\bar y)+i\bar \beta \bar d H(\bar y)\right],\\
\psi_2^{(-)}=&A\left[-F(\bar y)-\bar \beta(i+\bar y)G(\bar y)+i\bar \beta \bar d H(\bar y)\right].
\end{align}
Similarly, for Dresselhaus upper band 
\begin{align}
\psi_1^{(+)}=&A\left[F(\bar y)-\bar \beta(i-\bar y)G(\bar y)-i\bar \beta \bar d H(\bar y)\right],\\
\psi_2^{(+)}=&A\left[F(\bar y)-\bar \beta(i+\bar y)G(\bar y)+i\bar \beta \bar d H(\bar y)\right],
\end{align}
where$A=\frac{1}{\sqrt{2d}}\left(\frac{md}{2\pi it\hbar}\right)e^{\frac{i}{\hbar}\mu t}$.

\section{Special functions}\label{section2}
In this section, we provide some useful integral formula for the computation of the kernel in Eq.~(\ref{wavef}).  The hypergeometric function $_1F_1(a;b;z)$ can be expressed in an integral form given by
\begin{align}\label{A6}
_1F_1\!(a;c;\!z)\!=\!\frac{\Gamma(1\!+\!a\!-\!c)}{\Gamma(a)}\frac{(c\!-\!1)!}{2\pi i}\!\!\oint\!\mathrm ds(s\!-\!1)^{c-a-1}\!s^{a-\!1}e^{sz}\!,
\end{align}
where the integral contour is around the pole $s=1$ and c is an integer.  It is involved in the integration
\begin{align}\label{A2}
\frac{n!}{2a^{n+1}}\ _1F_1(n\!+\!1;1;-\frac{b^2}{4a})\!=\!\int_0^{\infty}\!\!e^{-ay^2}\!J_0(by)y^{2n+1}\mathrm dy,
\end{align}
where
\begin{align}\label{A1}
J_0(\frac{\overline r p}{\hbar})=\frac{1}{2\pi}\int_0^{2\pi}\mathrm d\phi e^{\frac{i}{\hbar}p(\overline x \cos\phi+\overline y \sin\phi)}
\end{align}
is the Bessel function of the first kind, and $\bar r=\sqrt{\bar x^2+\bar y^2}$.  In the following, we list some formula used in the computation.
\begin{align}\label{A3}
\int_0^{2\pi}\mathrm d\phi\cos \phi e^{\frac{i}{\hbar}p(\overline x \cos\phi+\overline y \sin\phi)}=-i2\pi \frac{\bar x}{\bar r}J_0'(\frac{\overline r p}{\hbar})
\end{align}
\begin{align}\label{A4}
\int_0^{2\pi}\mathrm d\phi\sin \phi e^{\frac{i}{\hbar}p(\overline x \cos\phi+\overline y \sin\phi)}=-i2\pi \frac{\bar y}{\bar r}J_0'(\frac{\overline r p}{\hbar})
\end{align}
\begin{align}\label{A5}
\int^{\infty}_0\!\!e^{-ay^2}\!J_1(by)y^{2n+2}\mathrm dy\!=\!\frac{b(n\!+\!1)!}{4a^{n+2}}\ _1F_1(n\!+\!2;2;-\frac{b^2}{4a})
\end{align}

%

\end{document}